\documentclass{emulateapj}
\usepackage{graphicx}
\usepackage{natbib}
\usepackage{aalongtable}

\slugcomment{Astronomical Journal in press}

\shorttitle{Chemical Diversity in High-Mass Star Formation}
\shortauthors{Beuther et al.}

\begin{document}

%% LaTeX will automatically break titles if they run longer than
%% one line. However, you may use \\ to force a line break if
%% you desire.

\title{Chemical Diversity in High-Mass Star Formation}

%% Use \author, \affil, and the \and command to format
%% author and affiliation information.
%% Note that \email has replaced the old \authoremail command
%% from AASTeX v4.0. You can use \email to mark an email address
%% anywhere in the paper, not just in the front matter.
%% As in the title, use \\ to force line breaks.

\author{H.~Beuther\altaffilmark{1}, Q.~Zhang\altaffilmark{2},
  E.A.~Bergin\altaffilmark{3}, and T.K. Sridharan\altaffilmark{2}}
\altaffiltext{1}{Max-Planck-Institute for Astronomy, K\"onigstuhl 17,
  69117 Heidelberg, Germany}
\altaffiltext{2}{Harvard-Smithsonian Center for Astrophysics, 60 Garden Street,
              Cambridge, MA 02138, USA}
\altaffiltext{3}{University of Michigan, 825 Dennison Building, 
             500 Church Street, Ann Arbor, MI 48109-1042}
%\affil{Astronomy Department,
%  University of California, Berkeley, CA 94720}

%\author{C. D. Biemesderfer\altaffilmark{4,5}}
%\affil{National Optical Astronomy Observatories, Tucson, AZ 85719}
%\email{aastex-help@aas.org}

%\and

%\author{R. J. Hanisch\altaffilmark{5}}
%\affil{Space Telescope Science Institute, Baltimore, MD 21218}

%% Notice that each of these authors has alternate affiliations, which
%% are identified by the \altaffilmark after each name.  Specify alternate
%% affiliation information with \altaffiltext, with one command per each
%% affiliation.

%\altaffiltext{1}{Visiting Astronomer, Cerro Tololo Inter-American Observatory.
%CTIO is operated by AURA, Inc.\ under contract to the National Science
%Foundation.}
%\altaffiltext{2}{Society of Fellows, Harvard University.}
%\altaffiltext{3}{present address: Center for Astrophysics,
%    60 Garden Street, Cambridge, MA 02138}
%\altaffiltext{4}{Visiting Programmer, Space Telescope Science Institute}
%\altaffiltext{5}{Patron, Alonso's Bar and Grill}

%% Mark off your abstract in the ``abstract'' environment. In the manuscript
%% style, abstract will output a Received/Accepted line after the
%% title and affiliation information. No date will appear since the author
%% does not have this information. The dates will be filled in by the
%% editorial office after submission.

\begin{abstract}
  Massive star formation exhibits an extremely rich chemistry.
  However, not much evolutionary details are known yet, especially at
  high spatial resolution. Therefore, we synthesize previously
  published Submillimeter Array high-spatial-resolution spectral line
  observations toward four regions of high-mass star formation that
  are in various evolutionary stages with a range of luminosities.
  Estimating column densities and comparing the spatially resolved
  molecular emission allows us to characterize the chemical evolution
  in more detail.  Furthermore, we model the chemical evolution of
  massive warm molecular cores to be directly compared with the data.
  The four regions reveal many different characteristics.  While some
  of them, e.g., the detection rate of CH$_3$OH, can be explained by
  variations of the average gas temperatures, other features are
  attributed to chemical effects. For example, C$^{34}$S is observed
  mainly at the core-edges and not toward their centers because of
  temperature-selective desorption and successive gas-phase chemistry
  reactions. Most nitrogen-bearing molecules are only found toward the
  hot molecular cores and not the earlier evolutionary stages,
  indicating that the formation and excitation of such complex
  nitrogen-bearing molecules needs significant heating and time to be
  fully developed.  Furthermore, we discuss the observational
  difficulties to study massive accretion disks in the young deeply
  embedded phase of massive star formation.  The general potential and
  limitations of such kind of dataset are discussed, and future
  directions are outlined.  The analysis and modeling of this source
  sample reveals many interesting features toward a chemical
  evolutionary sequence.  However, it is only an early step, and many
  observational and theoretical challenges in that field lie ahead.
\end{abstract}

%% Keywords should appear after the \end{abstract} command. The uncommented
%% example has been keyed in ApJ style. See the instructions to authors
%% for the journal to which you are submitting your paper to determine
%% what keyword punctuation is appropriate.


%% From the front matter, we move on to the body of the paper.
%% In the first two sections, notice the use of the natbib \citep
%% and \citet commands to identify citations.  The citations are
%% tied to the reference list via symbolic KEYs. The KEY corresponds
%% to the KEY in the \bibitem in the reference list below. We have
%% chosen the first three characters of the first author's name plus
%% the last two numeral of the year of publication as our KEY for
%% each reference.

%% Authors who wish to have the most important objects in their paper
%% linked in the electronic edition to a data center may do so by tagging
%% their objects with \objectname{} or \object{}.  Each macro takes the
%% object name as its required argument. The optional, square-bracket 
%% argument should be used in cases where the data center identification
%% differs from what is to be printed in the paper.  The text appearing 
%% in curly braces is what will appear in print in the published paper. 
%% If the object name is recognized by the data centers, it will be linked
%% in the electronic edition to the object data available at the data centers  
%%
%% Note that for sources with brackets in their names, e.g. [WEG2004] 14h-090,
%% the brackets must be escaped with backslashes when used in the first
%% square-bracket argument, for instance, \object[\[WEG2004\] 14h-090]{90}).
%%  Otherwise, LaTeX will issue an error. 

\section{Introduction}
\label{intro}

One of the main interests in high-mass star formation research today
is a thorough characterization of the expected massive accretion disks
(e.g., \citealt{yorke2002,krumholz2006b}). Observations of this type
require high spatial resolution, and disks have been suggested in
numerous systems, although not always with the same tracers (for a
recent compilation see \citealt{cesaroni2006}). A major complication
is the large degree of chemical diversity in these regions.  This is
seen on large scales in terms of the large number of molecular
detections (e.g., \citealt{schilke1997b,vandishoeck1998}) but in
particular on the small scale (of the order $10^4$\,AU and even
smaller) where significant chemical differentiation is found (e.g.,
\citealt{beuther2005a}).   At present, we are only beginning to
decipher the chemical structure of these objects at high spatial
resolution, whereas our understanding of the physical properties
better allows us to place regions into an evolutionary context (e.g.,
\citealt{beuther2006b,zinnecker2007}). It is not yet clear how our
emerging picture of the physical evolution is related to the observed
chemical diversity and evolution.

Various theory groups work on the chemical evolution during massive
star formation (e.g.,
\citealt{caselli1993,millar1997,charnley1997,viti2004,nomura2004,wakelam2005b,doty2002,doty2006}),
and the results are promising. However, the observational database to
test these models against is still relatively poor. Some single-dish
low-spatial-resolution line surveys toward several sources do exist,
but they are all conducted with different spatial resolution and
covering different frequency bands (e.g.,
\citealt{blake1987,macdonald1996,schilke1997b,hatchell1998b,mccutcheon2000,vandertak2000,vandertak2003,johnstone2003,bisschop2007}).
Furthermore, the chemical structure in massive star-forming regions is
far from uniform, and at high resolution one observes spatial
variations between many species, prominent examples are Orion-KL,
W3OH/H$_2$O or Cepheus A (see, e.g.,
\citealt{wright1996,wyrowski1999,beuther2005a,brogan2007}).

Single-dish studies targeted larger source samples at low spatial
resolution and described the averaged chemical properties of the
target regions (e.g., \citealt{hatchell1998b,bisschop2007}). However,
no consistent chemical investigation of a sample of massive
star-forming regions exists at high spatial resolution.  To obtain an
observational census of the chemical evolution at high spatial
resolution and to build up a database for chemical evolutionary models
of massive star formation, it is important to establish a rather
uniformly selected sample of massive star-forming regions in various
evolutionary stages.  Furthermore, this sample should be observed in
the same spectral setup at high spatial resolution. While the former
is necessary for a reliable comparison, the latter is crucial to
disentangle the chemical spatial variations in the complex massive
star-forming regions.  Because submm interferometric imaging is a time
consuming task, it is impossible to observe a large sample in a short
time.  Hence, it is useful to employ synergy effects and observe
various sources over a few years in the same spectral lines.

We have undertaken such a chemical survey of massive molecular cores
containing high-mass protostars in different evolutionary stages using
the Submillimeter Array (SMA\footnote{The Submillimeter Array is a
  joint project between the Smithsonian Astrophysical Observatory and
  the Academia Sinica Institute of Astronomy and Astrophysics, and is
  funded by the Smithsonian Institution and the Academia Sinica.},
\citealt{ho2004}) since 2003 in exactly the same spectral setup. The
four massive star-forming regions span a range of evolutionary stages
and luminosities: (1) the prototypical hot molecular core (HMC)
Orion-KL (\citealt{beuther2004g,beuther2005a}), (2) an HMC at larger
distance G29.96 \citep{beuther2007d}, and two regions in a presumably
earlier evolutionary phase, namely (3) the younger but similar
luminous High-Mass Protostellar Object (HMPO) IRAS\,23151+5912
\citep{beuther2007f} and (4) the less luminous HMPO IRAS\,05358+3543
\citep{leurini2007}. Although the latter two regions also have central
temperatures $>$100\,K, qualifying them as ``hot'', their molecular
line emission is considerably weaker than from regions which are
usually termed HMCs. Therefore, we refer to them from now on as
early-HMPOs (see also the evolutionary sequence in
\citealt{beuther2006b}). Table \ref{source_parameters} lists the main
physical parameters of the selected target regions.

The SMA offers high spatial resolution (of the order $1''$) and a
large enough instantaneous bandwidth of 4\,GHz to sample numerous
molecular transitions simultaneously (e.g., $^{28}$SiO and its rarer
isotopologue $^{30}$SiO, a large series of CH$_3$OH lines in the
$v_t=0,1,2$ states, CH$_3$CN, HCOOCH$_3$, SO, SO$_2$, and many more
lines in the given setup, see \S\ref{data}). Each of the these
observations have been published separately where we provide detailed
discussions of the particularities of each source
\citep{beuther2004g,beuther2005a,beuther2007f,leurini2007}). These
objects span an evolutionary range where the molecular gas and ice
coated grains in close proximity to the forming star are subject
to increasing degrees of heating.  In this fashion, volatiles will be
released from ices near the most evolved (luminous) sources altering
the surrounding gas-phase chemical equilibrium and molecular emission.
In this paper, we synthesize these data and effectively re-observe
these systems at identical physical resolution in order to identify
coherent trends. Our main goals are to search for (a) trends in the
chemistry as a function of evolutionary state and (b) to explore, in a
unbiased manner, the capability of molecular emission to trace
coherent velocity structures that have, in the past, been attribute to
Keplerian disks.  We will demonstrate that the ability of various
tracers to probe the innermost region, where a disk should reside,
changes as a function of evolution.

\begin{table*}[htb]
%\begin{center}
\caption{Source parameters}
\begin{tabular}{lrrrr}
\hline
\hline
                   & Orion-KL$^b$& G29.96$^b$   & 23151$^b$& 05358$^b$ \\
\hline
$L^a$\,[L$_{\odot}$] & $10^5$   & $9\times 10^4$  & $10^5$ & $10^{3.8}$ \\
$d$\,[pc]          & 450      & 6000            & 5700   & 1800 \\
$M_{\rm{gas}}^{a,c}$\,[M$_{\odot}$]& 140$^d$ & 2500$^e$ & 600 & 300 \\
$T_{\rm{rot}}^f$\,[K] & 300 & 340 & 150 & 220 \\
$N_{\rm{peak}}(\rm{H_2})^g$\,[cm${^-2}$]  & $9\times 10^{24}$ & $6\times 10^{24}$ & $2\times 10^{24}$ & $2\times 10^{24}$ \\
Type               & HMC      & HMC             & early-HMPO& early-HMPO \\
\hline
\hline
\end{tabular}
\footnotesize{~\\
  $^a$ Luminosities and masses are derived from single-dish data. Since most regions split up into multiple sources, individual values for sub-members are lower.\\
  $^b$ The SMA data are first published in \citealt{beuther2005a,beuther2007d,beuther2007f} and \citet{leurini2007}. Other parameters are taken from \citet{menten1995,olmi2003,sridha,beuther2002a}.\\
  $^c$ The integrated masses should be accurate within a factor 5 \citep{beuther2002a}.\\
  $^d$ This value was calculated from the 870\,$\mu$m flux of \citet{schilke1997b} following \citet{hildebrand1983} assuming an average temperature of 50\,K.\\
  $^e$ This value was calculated from the 850\,$\mu$m flux of \citet{thompson2006} following \citet{hildebrand1983} as in comment $b$.\\
  $^f$ Peak rotational temperatures derived from CH$_3$OH.\\
  $^g$ H$_2$ column densities toward the peak positions derived from the submm dust continuum observatios \citep{beuther2004g,beuther2007d,beuther2007f,beuther2007c}.}
\label{source_parameters}
%\end{center}
\end{table*}

\section{Data}
\label{data}

The four sources were observed with the SMA between 2003 and 2005 in
several array configurations achieving (sub)arcsecond spatial
resolution. For detailed observational description, see
\citet{beuther2005a,beuther2007d,beuther2007f} and
\citet{leurini2007}. The main point of interest to be mentioned here
is that all four regions were observed in exactly the same spectral
setup. The receivers operated in a double-sideband mode with an IF
band of 4-6\,GHz so that the upper and lower sideband were separated
by 10\,GHz. The central frequencies of the upper and lower sideband
were 348.2 and 338.2\,GHz, respectively.  The correlator had a
bandwidth of 2\,GHz and the channel spacing was 0.8125\,MHz, resulting
in a nominal spectral resolution of $\sim$0.7\,km\,s$^{-1}$. However,
for the analysis presented below, we smoothed the data-cubes to
2\,km\,s$^{-1}$. The spatial resolution of the several datasets is
given in Table \ref{resolution}. Line identifications were done in an
iterative way: we first compared our data with the single-dish line
survey of Orion by \citet{schilke1997b} and then refined the analysis
via the molecular spectroscopy catalogs of JPL and the Cologne
database of molecular spectroscopy CDMS
\citep{poynter1985,mueller2002}.

\begin{table}[htb]
\caption{Spatial resolution}
\begin{tabular}{lrrrr}
\hline
\hline
             & Orion-KL & G29.96 & 23151 & 05358 \\
\hline
Cont. [$''$]   & $0.8\times 0.7$ & $0.4\times 0.3$ & $0.6\times 0.5$ & $1.1\times 0.6$ \\
Av. Cont. [AU] & 340 & 2100 & 3100 & 1500 \\
Line  [$''$]   & $1.4\times 1.1 $& $0.6\times 0.5$ & $1.1\times 0.8$ & $1.4\times 0.8$\\
Av. Line  [AU] & 560 & 3300 & 5400 & 2000 \\
\hline
\hline
\end{tabular}
\label{resolution}
\end{table}

\section{Results and Discussion}

\subsection{Submm continuum emission}

\begin{figure*}[htb]
\includegraphics[angle=-90,width=\textwidth]{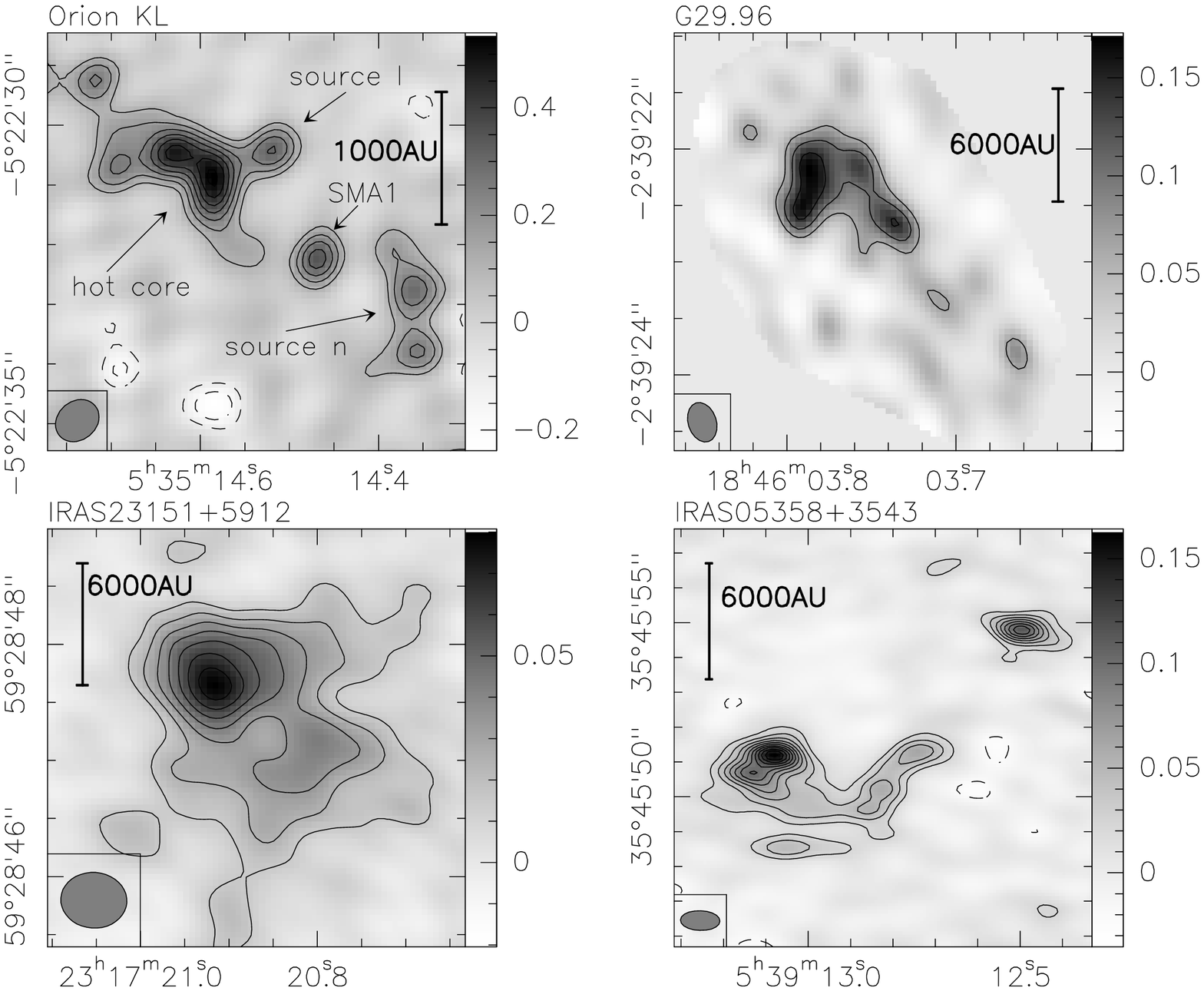}
\caption{862\,$\mu$m continuum images toward the four massive
  star-forming regions. Positive and negative features are shown as
  solid and dashed contours, respectively. The contouring always
  starts at the $3\sigma$ levels and continues in $2\sigma$ steps
  (with $1\sigma$ values of 35, 21, 5, 7\,mJy\,beam$^{-1}$,
  respectively). These figures are adaptions from the papers by
  \citet{beuther2005a,beuther2007d,beuther2007f,beuther2007c}.
  Scale-bars are presented in each panel, and the wedges show the flux
  density scales in Jy. The axis are labeled in R.A.  (J2000) and Dec.
  (J2000).  The synthesized beams sizes are listed in Table
  \ref{resolution}.}
\label{cont_sample}
\end{figure*}

To set the four regions spatially into context, Figure
\ref{cont_sample} presents the four submm continuum images obtained
with the SMA. As one expects in a clustered mode of massive star
formation, all regions exhibit multiple structures with the number of
sub-sources $\geq 2$. As discussed in the corresponding papers, while
most of the submm continuum peaks likely correspond to a
embedded protostellar sources, this is not necessarily the case for
all of them.  Some of the submm continuum peaks could be
externally heated gas clumps (e.g., the Orion hot core peak,
\citealt{beuther2004g}) or may be produced by shock interactions with
molecular outflows (e.g., IRAS\,05358+3543, \citealt{beuther2007c}).
In the following spectral line images, we will always show the
corresponding submm continuum map in grey-scale as a reference frame.

\subsection{Spectral characteristics as a function of evolution}

\begin{figure*}[htb]
\includegraphics[angle=-90,width=0.5\textwidth]{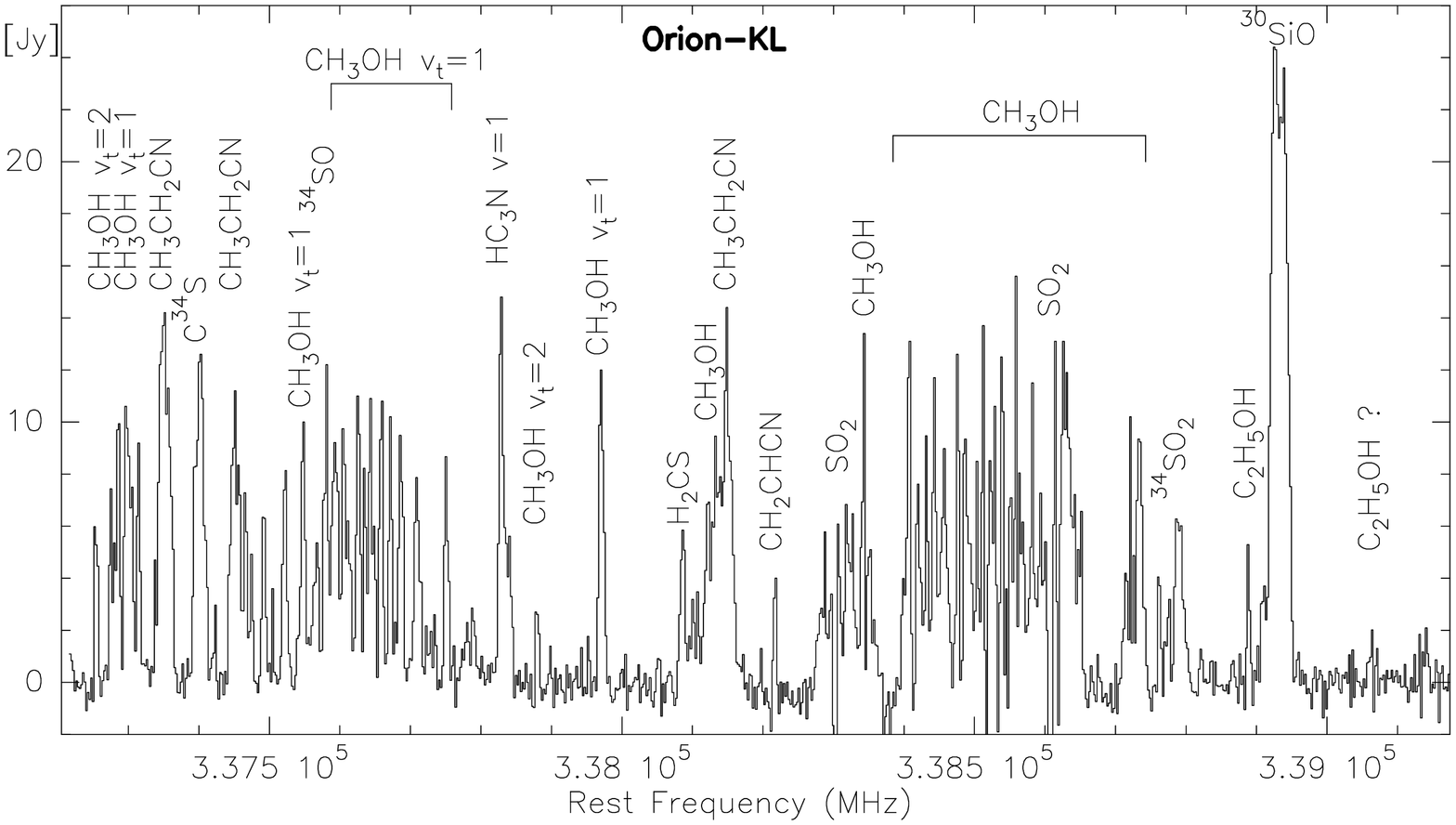}
\includegraphics[angle=-90,width=0.5\textwidth]{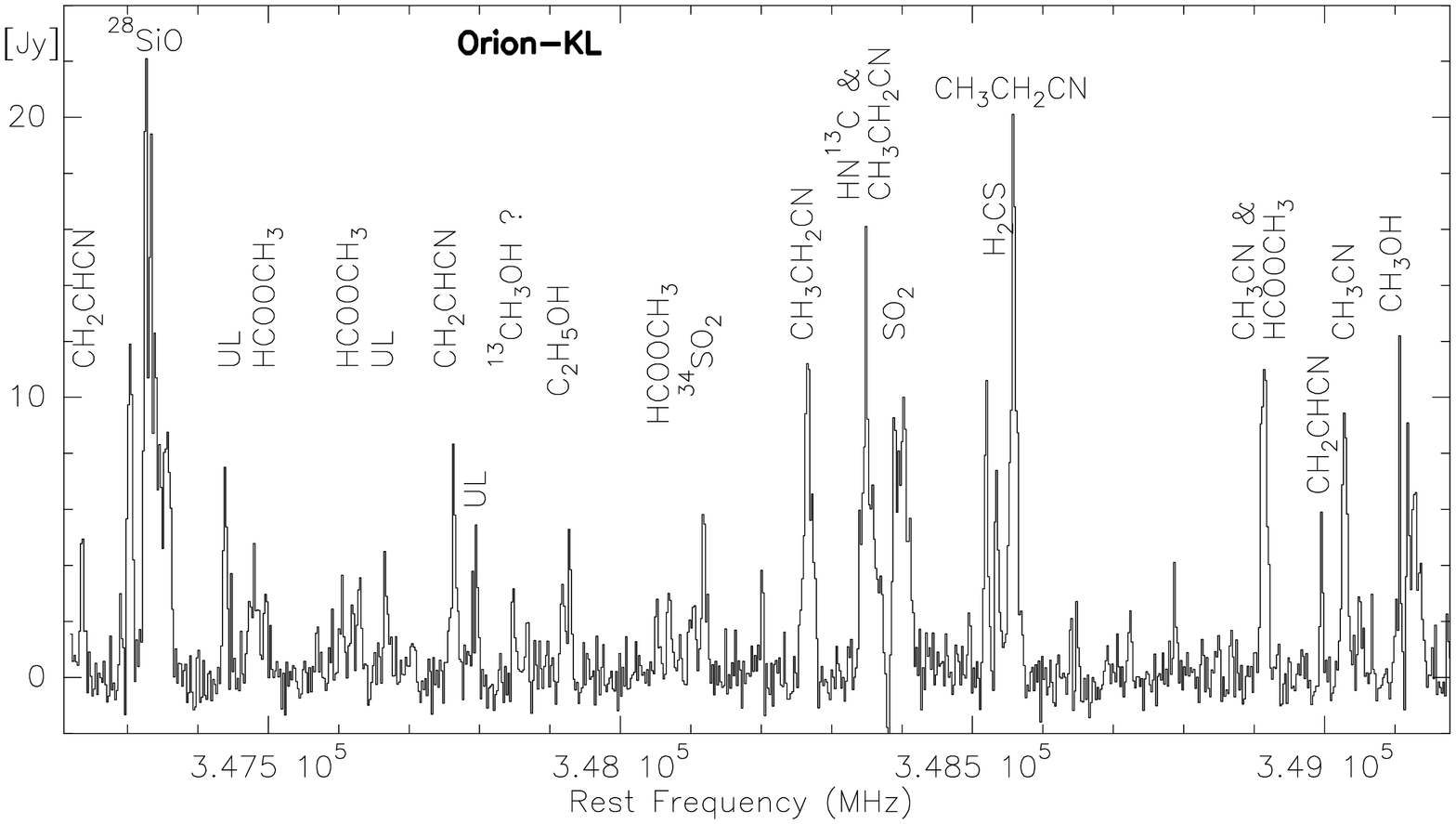}\\
\includegraphics[angle=-90,width=0.5\textwidth]{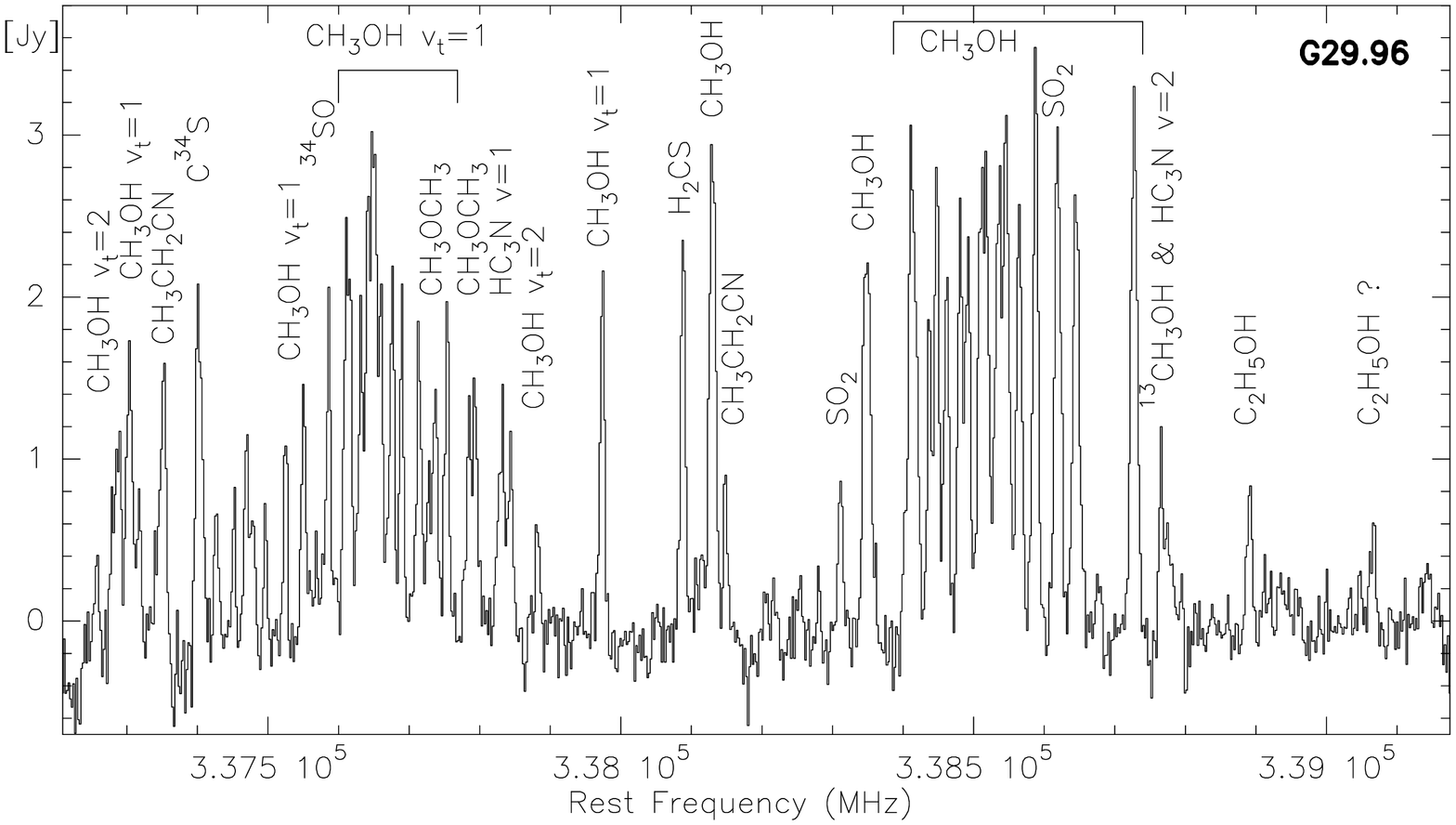}
\includegraphics[angle=-90,width=0.5\textwidth]{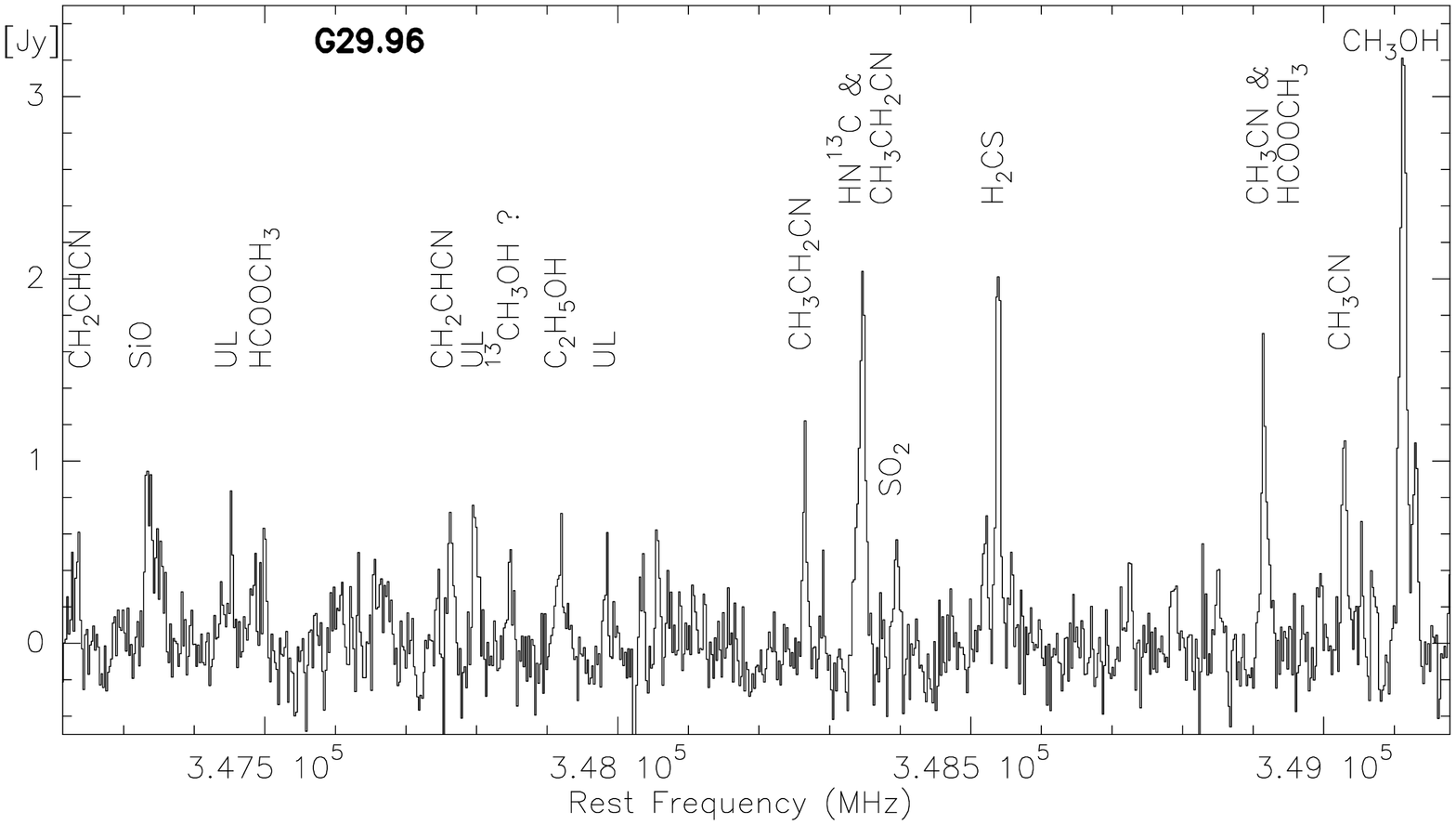}\\
\includegraphics[angle=-90,width=0.5\textwidth]{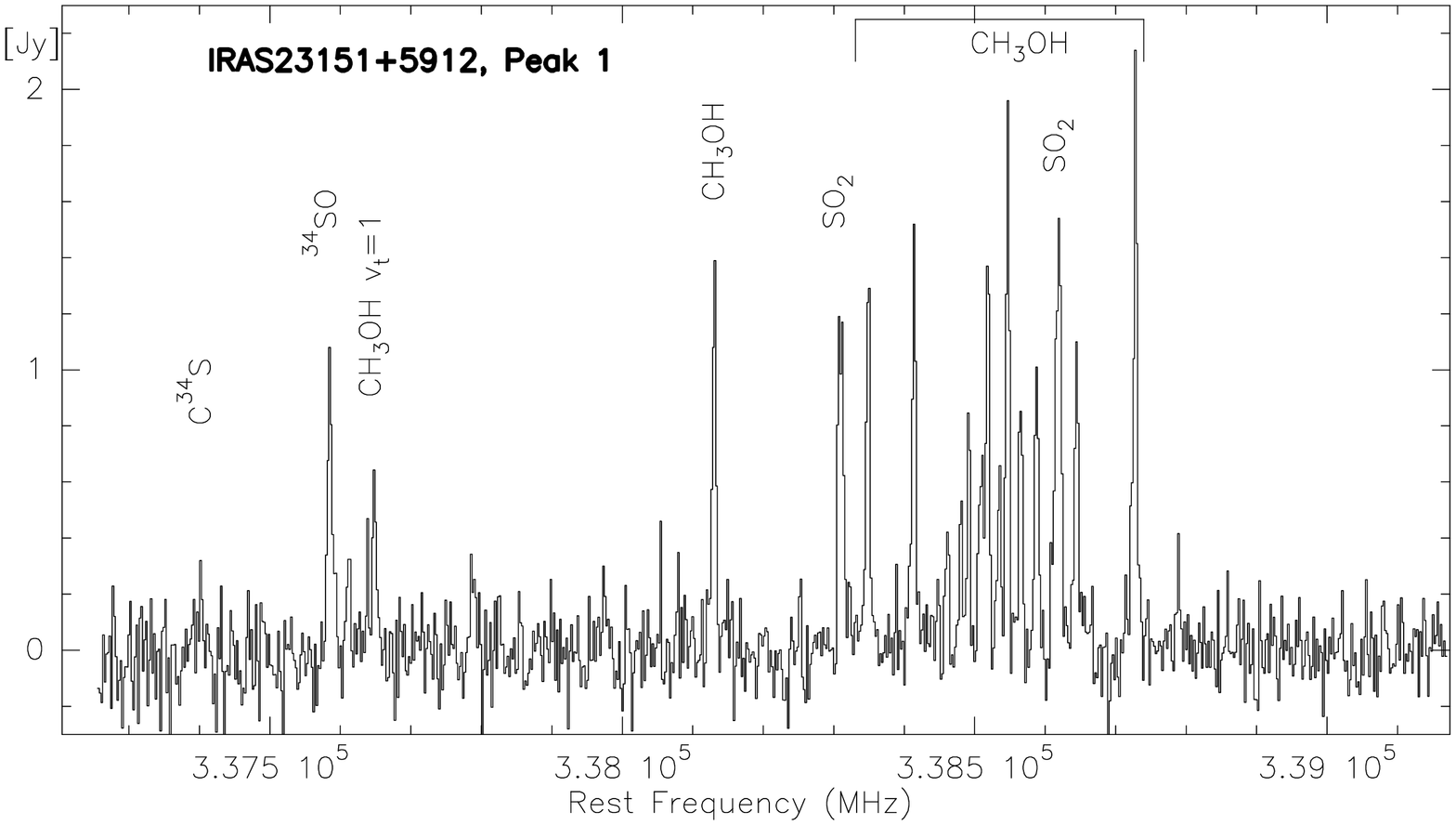}
\includegraphics[angle=-90,width=0.5\textwidth]{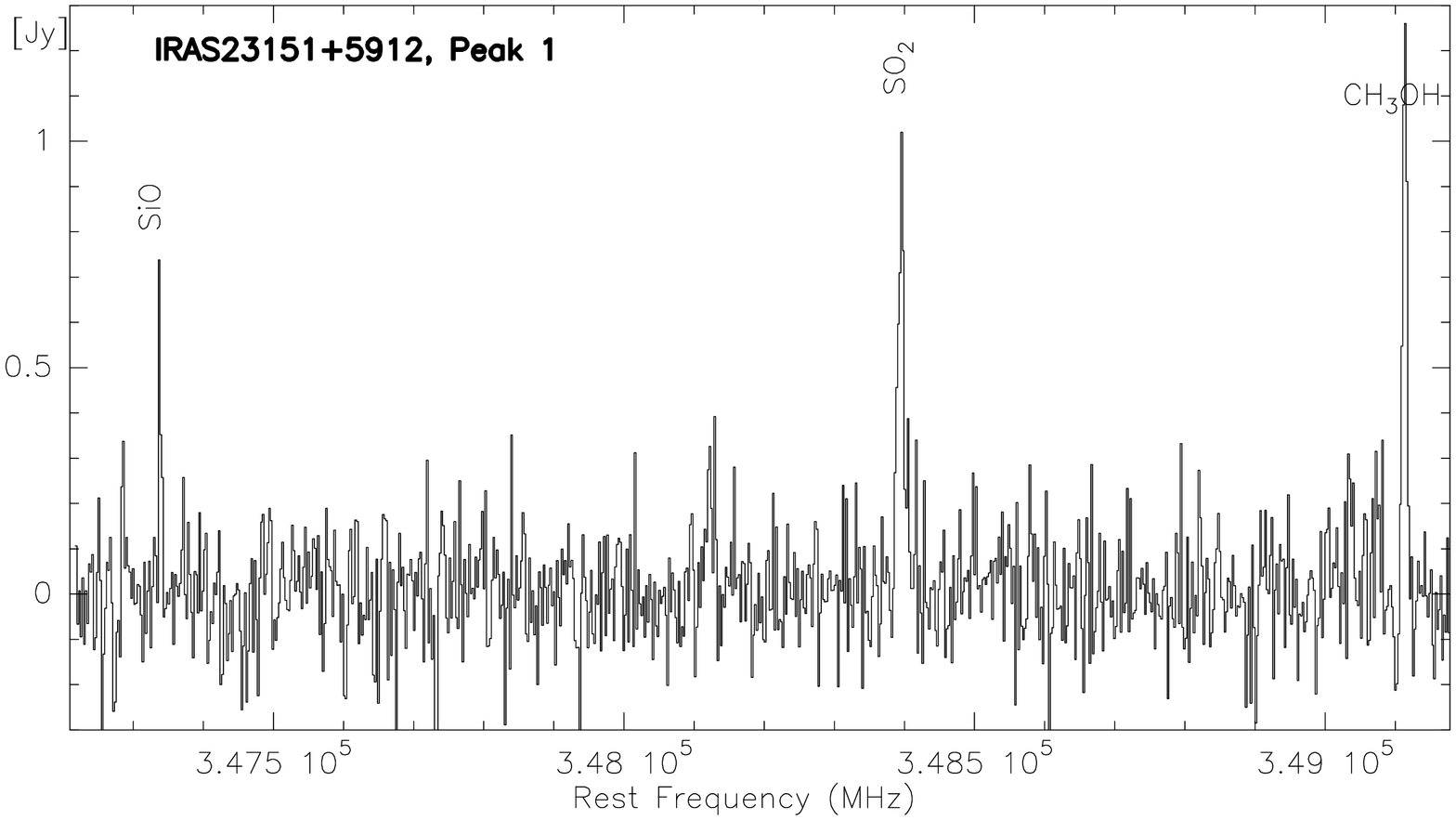}\\
\includegraphics[angle=-90,width=0.5\textwidth]{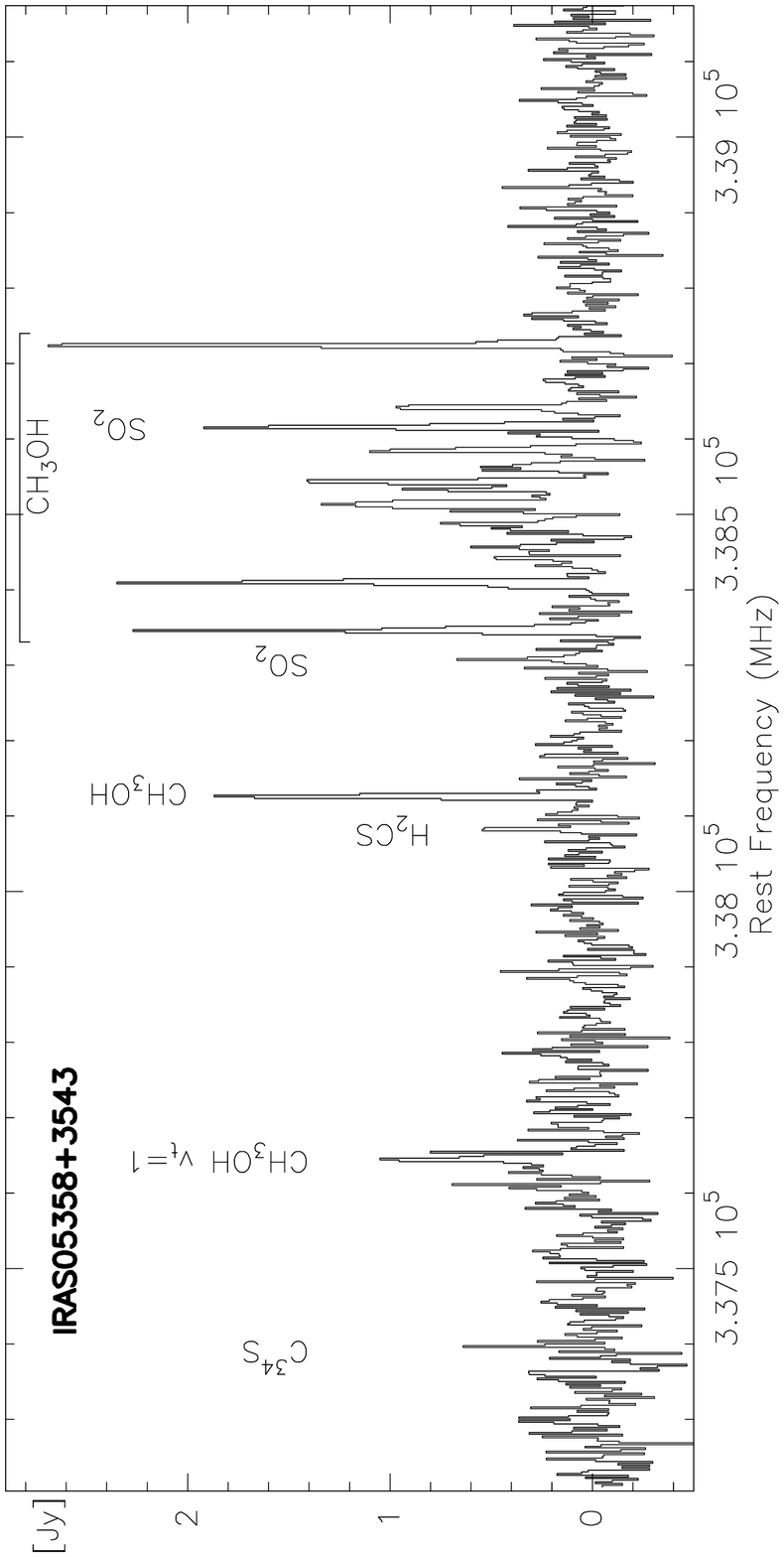}
\includegraphics[angle=-90,width=0.5\textwidth]{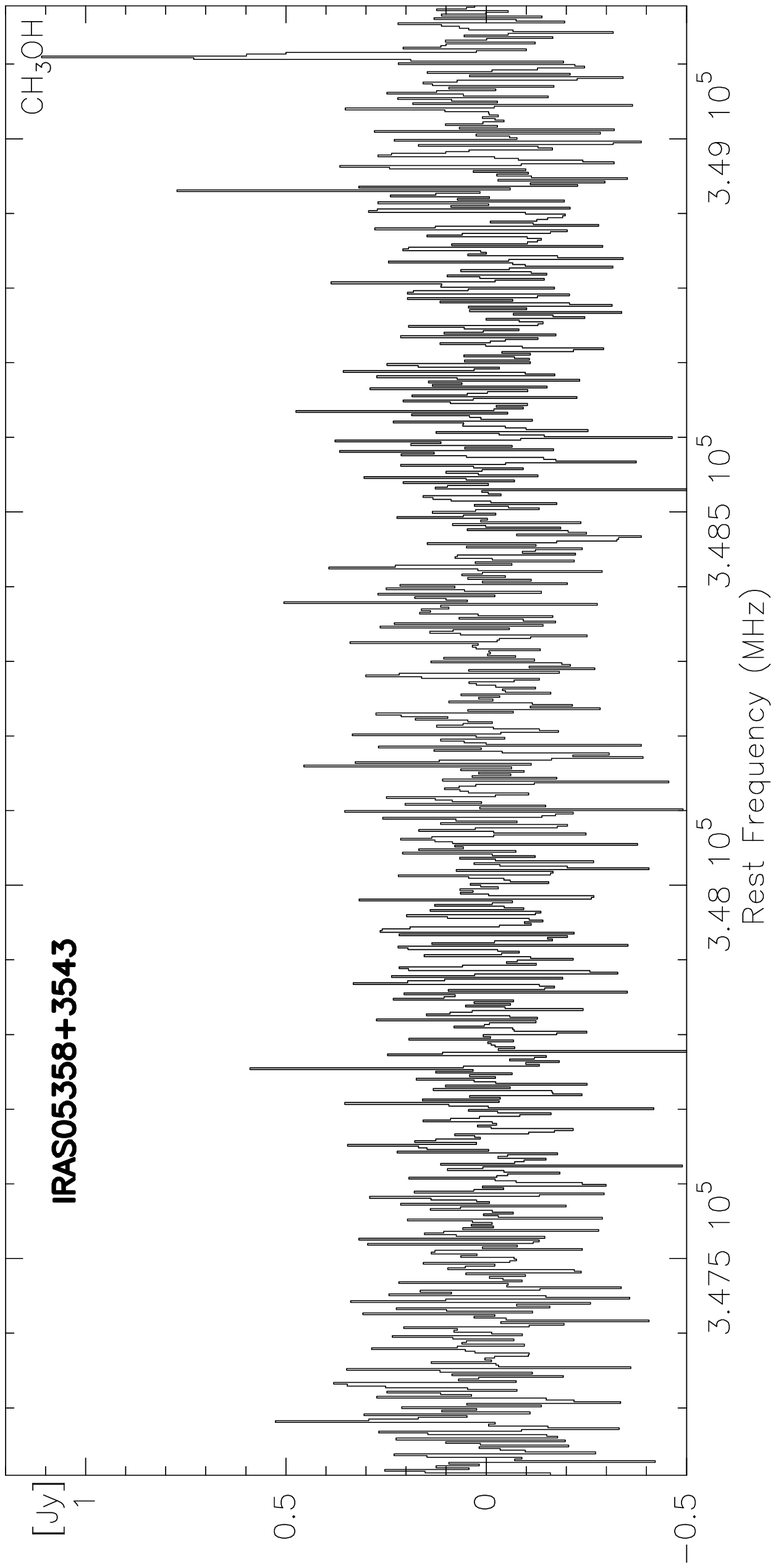}
\caption{SMA spectra extracted from the final data-cubes in the image
  domain toward four massive star-forming regions (Orion-KL top-row,
  G29.96 second row, IRAS\,23151+5912 third row and IRAS\,05358+3543
  fourth row). For a better comparison, the data-cubes were all
  smoothed to the same spatial resolution of $\sim$5700\,AU from the
  current IRAS\,23151+3543 dataset. The circles in
  Fig.~\ref{ch3oh_sample} outline the corresponding spatial regions.
  The spectral resolution in all spectra is 2\,km/s.}
\label{sample_spectra}
\end{figure*}

Because the four regions are at different distances (Table
\ref{source_parameters}), to compare the overall spectra we smoothed
all datasets to the same linear spatial resolution of $\sim$5700\,AU.
Figure \ref{sample_spectra} presents the final spectra extracted at
this common resolution toward the peak positions of all four regions.
Furthermore, we present images at the original spatial resolution of
the four molecular species or vibrationally-torsionally excited lines
that are detected toward all four target regions (Figs.~3 to 6).

The spectral characteristics between the HMCs and the early-HMPOs vary
considerably. Table \ref{linelistall} lists all detected lines in the
four regions with their upper energy levels $E_u/k$ and peak
intensities $S_{\rm{peak}}$ at a spatial resolution of $\sim$5700\,AU.

\subsubsection{Excitation and optical depth effects?}
\label{excitation_effects}

Since the line detections and intensities are not only affected by the
chemistry but also by excitation effects, we have to estimate
quantitatively how much the latter can influence our data. In local
thermodynamic equilibrium, the line intensities are to first order
depending on the Boltzmann factor $e^{\frac{-E_u}{kT}}$ and the
partition function $Q$:

$$\int I(T) dv \propto \frac{e^{\frac{-E_u}{kT}}}{Q(t)}$$ 

where $\frac{E_u}{k}$ is the upper level energy state. For polyatomic
molecules in the high-temperature limit, one can approximate
$Q(T)\propto \sqrt{T^3}$ (e.g., \citealt{blake1987}). To get a feeling
how much temperature changes affect lines with different
$\frac{E_u}{k}$, one can form the ratio of $\int I(T) dv$ at two
different temperatures:

$$\frac{\int I(2T) dv}{\int I(T) dv} = \frac{\sqrt{e^{\frac{E_u}{kT}}}}{\sqrt{2^3}}$$

Equating this ratio now for a few respective upper level energy states
and gas temperatures of $T=100$ \& 50\,K (Table \ref{excite}), we find
that the induced intensity changes mostly barely exceed a factor 2.
Only for very highly excited lines like
CH$_3$OH$(7_{1,7}-6_{1,6})v_t=1$ at relatively low temperatures
($T=50$\,K) do excitation effects become significant. However, at such
low temperatures, these highly excited lines emit well below our
detection limits, hence this case is not important for the present
comparison.  Therefore, excitation plays only a minor role in
producing the molecular line differences discussed below, and other
effects like the chemistry turn out to be far more important.

\begin{table}[htb]
\caption{Excitation effects}
\begin{tabular}{lrrr}
\hline
\hline
Line           & $\frac{E_u}{k}$ & $\frac{\int I(2T) dv}{\int I(T) dv}$ & $\frac{\int I(2T) dv}{\int I(T) dv}$\\
               & (K)             & @ $T$=100K & @ $T$=50K\\
\hline
C$^{34}$S(7--6) & 65 & 0.49 & 0.68 \\
SO$_2(18_{4,1}-18_{3,1})$ & 197 & 0.95 & 2.5 \\
CH$_3$OH$(7_{1,7}-6_{1,6})v_t=1$& 356 & 2.1 & 12.4 \\
~\\
\hline
\hline
\end{tabular}
\label{excite}
\end{table}

As shown in Table \ref{linelistall}, our spectral setup barely
contains few lines from rarer isotopologues. We therefore cannot
readily determine the optical depth of the molecular lines. While it
is likely that, for example, the ground state CH$_3$OH lines have
significant optical depths, rarer species and vibrationally excited
lines should be more optically thin. We checked this for a few
respective species (e.g., C$^{34}$S or SO$_2$) via running
large-velocity gradient models (LVG, \citealt{vandertak2007}) with
typical parameters for these kind of regions, confirming the overall
validity of optically thin emission for most lines (see
\S\ref{column_para} \& \S\ref{c34s}).  However, without additional
data, we cannot address this issue in more detail.

\subsubsection{Column densities}
\label{column_para}

Estimating reliable molecular column densities and/or abundances is a
relatively difficult task for interferometric datasets like those
presented here. The data are suffering from missing short spacings and
filter out large fractions of the gas and dust emission. Because of
the different nature of the sources and their varying distances, the
spatial filtering affects each dataset in a different way.
Furthermore, because of spatial variations between the molecular gas
distributions and the dust emission representing the H$_2$ column
densities, the spatial filtering affects the dust continuum and the
spectral line emission differently. On top of this, Figures
\ref{ch3oh_sample} to \ref{so2_sample} show that is several cases the
molecular line and dust continuum emission are even spatially offset,
preventing the estimation of reliable abundances.

While these problems make direct abundance estimates relative to H$_2$
impossible, nevertheless, we are at least able to estimate approximate
molecular column densities for the sources. Since we are dealing with
high-density regions, we can derive the column densities from the
spectra shown in Figure \ref{sample_spectra} assuming local
thermodynamic equilibrium (LTE) and optically thin emission. We
modeled the molecular emission of each species separately using the
XCLASS superset to the CLASS software developed by Peter Schilke
(priv.~comm.). This software package uses the line catalogs from JPL
and CDMS \citep{poynter1985,mueller2001}. The main free parameters for
the molecular spectra are temperature, source size and column density.
We used the temperatures given in Table \ref{source_parameters}
(except of Orion-KL where we used 200\,K because of the large
smoothing to 5700\,AU) with approximate source sizes estimated from
the dust continuum and spectral line maps. Then we produced model
spectra with the column density as the remaining free parameter.
Considering the missing flux and the uncertainties for temperatures
and source sizes, the derived column densities should be taken with
caution and only be considered as order-of-magnitude estimates. Table
\ref{column} presents the results for all sources and detected
molecules.

\subsubsection{General differences}

An obvious difference between the four regions is the large line
forest observed toward the two HMCs Orion-KL and G29.96 and the
progressively less detected molecular lines toward IRAS\,23151+5912
and IRAS\,05358+3543. Especially prominent is the difference in
vibrationally-torsionally excited CH$_3$OH lines: we detect many
transitions in the HMCs and only a single one in the two early-HMPOs.
Since the vibrationally-torsionally excited CH$_3$OH lines have higher
excitation levels $E_u/k$, this can be relatively easily explained by
on average lower temperatures of the molecular gas in the early-HMPOs.
Assuming an evolutionary sequence, we anticipate that the two
early-HMPOs will eventually develop similar line forests like the two
HMCs.

\begin{figure*}[htb]
\includegraphics[angle=-90,width=\textwidth]{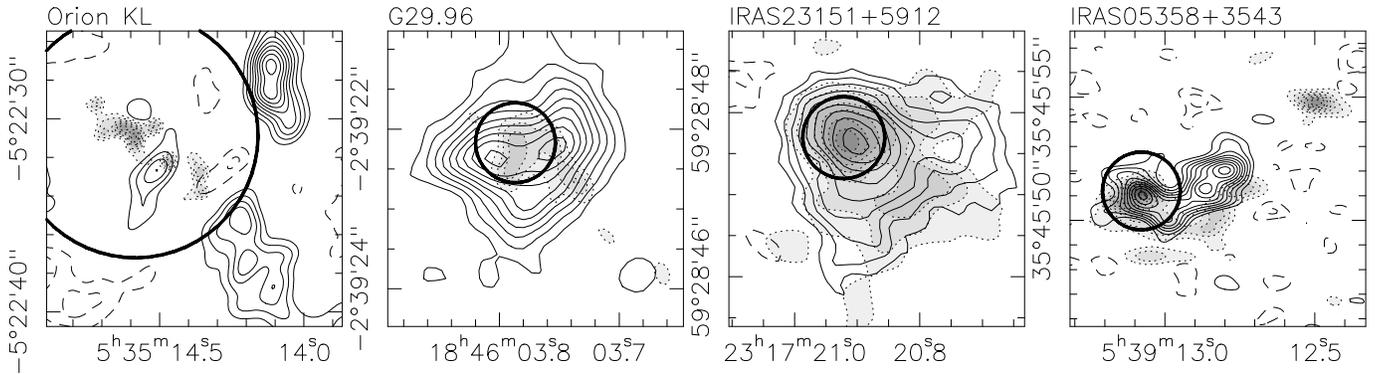}
\caption{CH$_3$OH contour images (line blend between
  CH$_3$OH$(7_{2,5}-6_{2,4})$ and CH$_3$OH$(7_{2,6}-6_{2,5})$) toward
  the four massive star-forming regions.  Positive and negative
  features are shown as solid and dashed contours. The contouring is
  done from 15 to 95\% (step 10\%) of the peak emission, and the peak
  emission values are 7.5, 0.9, 1.0 and 0.6\,Jy\,beam$^{-1}$ from left
  to right, respectively. The integration regimes for the four sources
  are [5,15], [90,104], [-60,-52] and [-20,-12]\,km\,s$^{-1}$. The
  grey-scale with dotted contours shows the submm continuum emission
  from Fig.~\ref{cont_sample}.  This figure is an adaption from the
  papers by
  \citet{beuther2005a,beuther2007d,beuther2007f,leurini2007}. The axis
  are labeled in R.A.  (J2000) and Dec. (J2000).  The spatial
  resolution is listed in Table \ref{resolution}. The circles
  represent the regions of diameter 5700\,AU used for the comparison
  spectra in Fig.~\ref{sample_spectra}.}
\label{ch3oh_sample}
\end{figure*}

\begin{figure*}[htb]
\includegraphics[angle=-90,width=\textwidth]{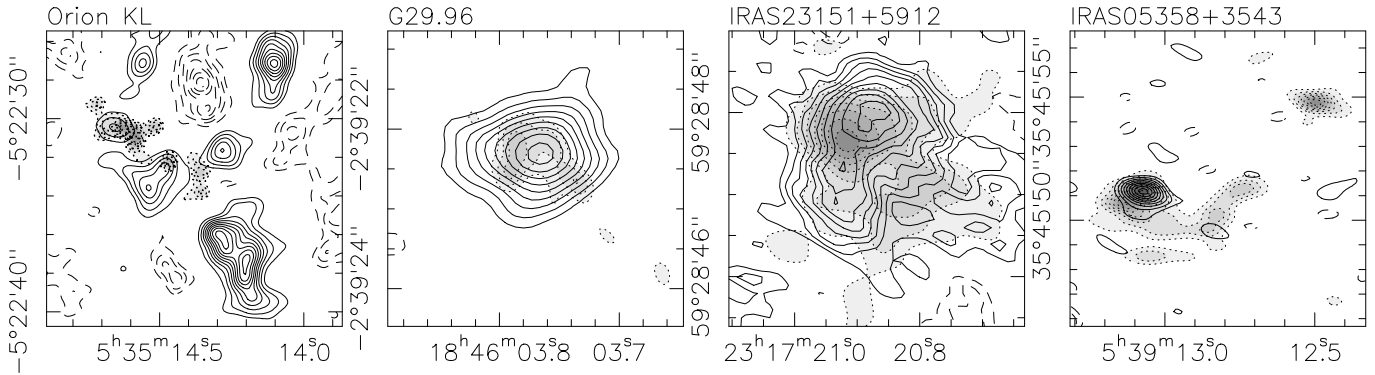}
\caption{Rotationally-torsionally excited
  CH$_3$OH$(7_{1,7}-6_{1,6})(v_t=1)$ images toward the four massive
  star-forming regions.  Positive and negative features are shown as
  solid and dashed contours. The contouring is done from 15 to 95\%
  (step 10\%) of the peak emission, and the peak emission values are
  7.5, 0.9, 1.0 and 0.6\,Jy\,beam$^{-1}$ from left to right,
  respectively. The integration regimes for the four sources are
  [3,13], [91,105], [-58,-54] and [-18,-10]\,km\,s$^{-1}$. The
  grey-scale with dotted contours shows the submm continuum emission
  from Fig.~\ref{cont_sample}. This figure is an adaption from the
  papers by
  \citet{beuther2005a,beuther2007d,beuther2007f,leurini2007}. The axis
  are labeled in R.A. (J2000) and Dec. (J2000). The spatial resolution
  is listed in Table \ref{resolution}.}
\label{ch3oh_vt1_sample}
\end{figure*}

Analyzing the spatial distribution of CH$_3$OH we find that it is
associated with several physical entities (Figs.~\ref{ch3oh_sample} \&
\ref{ch3oh_vt1_sample}). While it shows strong emission toward most
submm continuum peaks, it exhibits additional interesting features.
For example, the double-peaked structure in G29.96
(Fig~\ref{ch3oh_sample}) may be caused by high optical depth of the
molecular emission, whereas the lower optical depth
vibrationally-torsionally excited lines do peak toward the central
dust and gas core (Fig~\ref{ch3oh_vt1_sample}, smoothing the submm
continuum map to the lower spatial resolution of the line data, the
four submm sources merge into one central peak). In contrast, toward
Orion-KL the strongest CH$_3$OH features in the ground state and the
vibrationally-torsionally excited states are toward the south-western
region called the compact ridge. This is the interface between a
molecular outflow and the ambient gas.  Our data confirm previous work
which suggests an abundance enrichment (e.g., \citealt{blake1987}).
For instance in quiescent gas in Orion the ratio of CH$_3$OH/C$^{34}$S
$<$\,20 \citep{bergin1997}, while our data find a ratio of 100 (see
Table \ref{column}).  This is believed to be caused by outflow shock
processes in the dense surrounding gas (e.g., \citealt{wright1996}).
Furthermore, as will be discussed in \S\ref{disks}, there exist
observational indications in two of the sources (IRAS\,23151+5912 and
IRAS\,05358+3543) that the vibrationally-torsionally excited CH$_3$OH
may be a suitable tracer of inner rotating disk or tori structures.

Toward G29.96, we detect most lines previously also observed toward
Orion-KL, a few exceptions are the $^{30}$SiO line, some of the
vibrationally-torsionally state $v_t=2$ CH$_3$OH lines, some N-bearing
molecular lines from larger molecules like CH$_3$CH$_2$CN or
CH$_3$CHCN, as well as a few $^{34}$SO$_2$ and HCOOCH$_3$ lines.
While for some of the weaker lines this difference may partly be
attributed to the larger distance of G29.96, for other lines such an
argument is unlikely to hold. For example, the CH$_3$CH$_2$CN line at
348.55\,GHz is stronger than the neighboring H$_2$CS line in Orion-KL
whereas it remains undetected in G29.96 compared to the strong H$_2$CS
line there. This is also reflected in the different abundance ratio of
CH$_3$CH$_2$CN/H$_2$CS which is more than an order of magnitude larger
in Orion-KL compared with G29.96 (see Table \ref{column}).  Therefore,
these differences are likely tracing true chemical variations between
sources. In contrast, the only lines observed toward G29.96 but not
detected toward Orion-KL are a few CH$_3$OCH$_3$ lines.

The main SiO isotopologue $^{28}$SiO(8--7) is detected in all sources
but IRAS\,05358+3543. This is relatively surprising because SiO(2--1)
is strong in this region \citep{beuther2002d}, and the upper level
energy of the $J=8-7$ of $\sim$75\,K does not seem that
extraordinarily high to produce a non-detection. For example, the
detected CH$_3$OH$(7_{1,7}-6_{1,6})v_t=1$ transition has an upper
level energy of 356\,K. This implies that IRAS\,05358+3543 does have
warm molecular gas close to the central sources. However, the
outflow-components traced by SiO are at on average lower temperatures
(probably of the order 30\,K, e.g., \citealt{cabrit1990}) which may be
the cause of the non-detection in IRAS\,05358+3543. Furthermore, the
critical density of the SiO(8--7) line is about two orders of
magnitude higher than that of the (2--1) transition. Hence, the
density structure of the core may cause the (8--7) non-detection in
IRAS\,05358+3543 as well.

While the rarer $^{30}$SiO(8--7) isotopologue is detected toward
Orion-KL with nearly comparable strength as the main isotopologue
(Fig.~\ref{sample_spectra} and \citealt{beuther2005a}), we do not
detect it at all in any of the other sources.

A little bit surprising, the H$_2$CS line at 338.081\,GHz is detected
toward Orion-KL, G29.96 as well as the lowest luminosity source
IRAS\,05358+3543, however, it remains undetected toward the more
luminous HMPO IRAS\,23151+5912.  We are currently lacking a good
explanation for this phenomenon because H$_2$CS is predicted by most
chemistry networks as a parent molecule to be found early in the
evolutionary sequence (e.g., \citealt{nomura2004}). The sulphur and
nitrogen chemistries are also peculiar in this sample, and we outline
some examples below.

\subsubsection{Sulphur chemistry}
\label{c34s}

\begin{figure*}[htb]
\includegraphics[angle=-90,width=\textwidth]{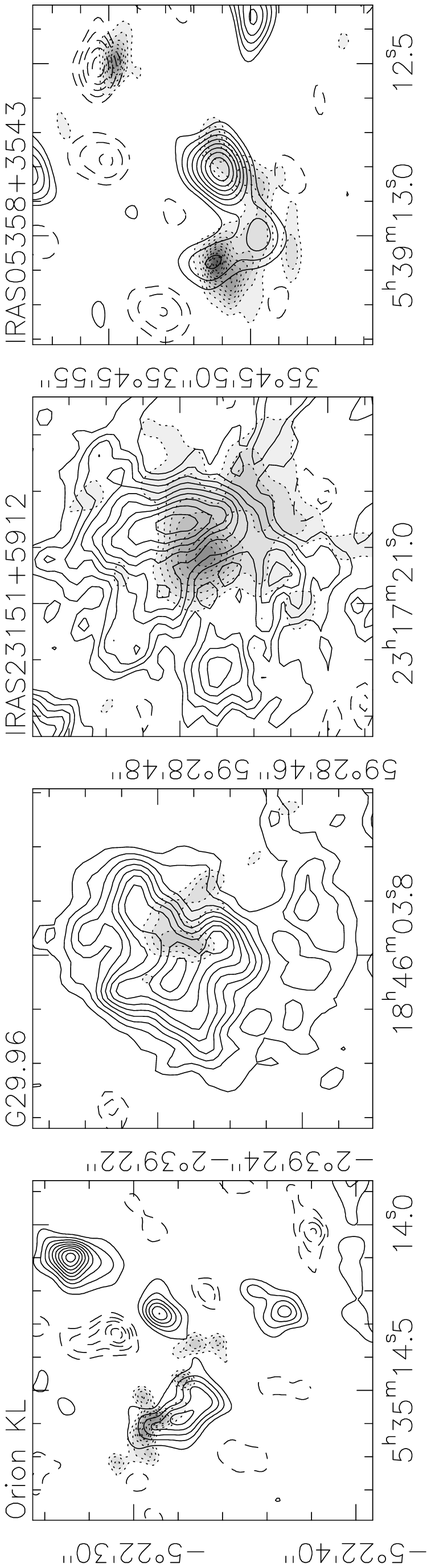}
\caption{C$^{34}$S(7--6) images toward the four massive star-forming
  regions.  Positive and negative features are shown as solid and
  dashed contours. The contouring is done from 15 to 95\% (step 10\%)
  of the peak emission, and the peak emission values are 7.5, 0.9, 1.0
  and 0.6\,Jy\,beam$^{-1}$ from left to right, respectively. The
  integration regimes for the four sources are [-2,14], [92,104],
  [-58,-51] and [-17,-13]\,km\,s$^{-1}$. The grey-scale with dotted
  contours shows the submm continuum emission from
  Fig.~\ref{cont_sample}. This figure is an adaption from the papers
  by \citet{beuther2005a,beuther2007d,beuther2007f,leurini2007}. The
  axis are labeled in R.A.  (J2000) and Dec.  (J2000).  The spatial
  resolution is listed in Table \ref{resolution}.}
\label{c34s_sample}
\end{figure*}

The rare Carbon-sulphur isotopologue C$^{34}$S is detected toward all
four regions (Fig.~\ref{sample_spectra}). However, as shown in
Fig.~\ref{c34s_sample} C$^{34}$S does not peak toward the main submm
continuum peaks but is offset at the edge of the core. In the cases of
G29.96 and IRAS\,23151+5912 the C$^{34}$S morphology appears to wrap
around the main submm continuum peaks. Toward IRAS\,05358+3543
C$^{34}$S is also weak toward the strongest submm peak (at the eastern
edge of the image) but shows the strongest C$^{34}$S emission features
offset from a secondary submm continuum source (in the middle of the
image). Toward Orion-KL, weak C$^{34}$S emission is detected in the
vicinity of the hot core peak, whereas we find strong C$^{34}$S
emission peaks offset from the dust continuum emission.

\begin{figure*}[htb]
\includegraphics[angle=-90,width=\textwidth]{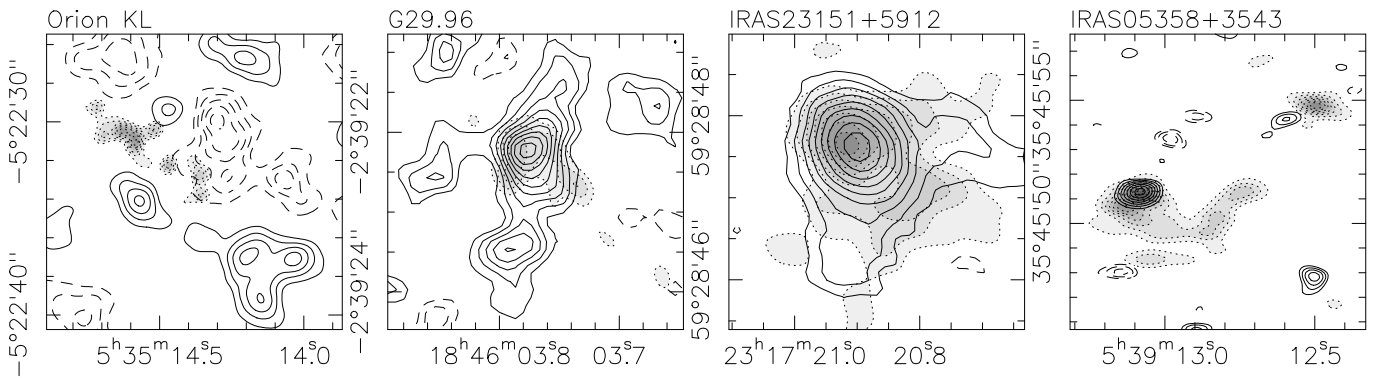}
\caption{SO$_2(14_{4,14}-18_{3,15})$ images toward the four massive
  star-forming regions.  Positive and negative features are shown as
  solid and dashed contours. The contouring is done from 15 to 95\%
  (step 10\%) of the peak emission, and the peak emission values are
  7.5, 0.9, 1.0 and 0.6\,Jy\,beam$^{-1}$ from left to right,
  respectively. Only for IRAS\,05358+1732 the contouring starts at the
  35\% level because of a worse signal-to-noise ratio. The integration
  regimes for the four sources are [0,20], [94,100], [-60,-50] and
  [-17,-13]\,km\,s$^{-1}$. The grey-scale with dotted contours shows
  the submm continuum emission from Fig.~\ref{cont_sample}. This
  figure is an adaption from the papers by
  \citet{beuther2005a,beuther2007d,beuther2007f,leurini2007}. The axis
  are labeled in R.A.  (J2000) and Dec. (J2000).  The spatial
  resolution is listed in Table \ref{resolution}.}
\label{so2_sample}
\end{figure*}

\begin{figure*}[htb]
\includegraphics[angle=-90,width=\textwidth]{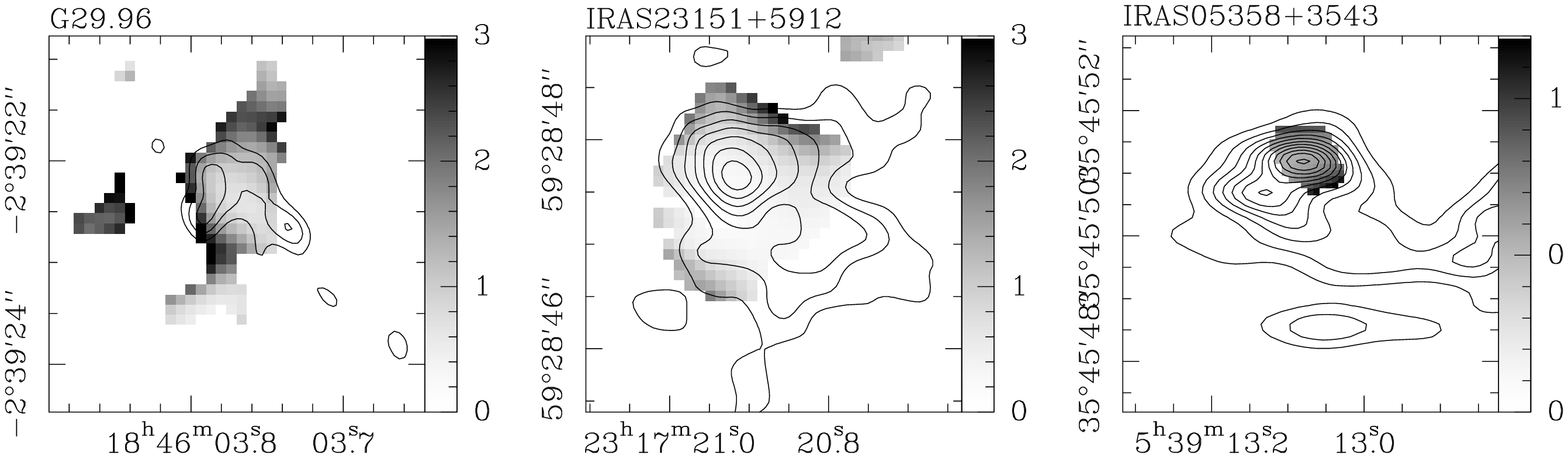}
\caption{Column density ratios between C$^{34}$S over SO$_2$ toward
  G29.96, IRAS\,23151+5912 and IRAS\,05358+3543 are shown in
  grey-scale. The contours present the corresponding submm continuum
  maps as in shown the previous figures. For IRAS\,05358+3543 we zoom
  only into the central region to better show the ratio variations.}
\label{ratio}
\end{figure*}

To check whether our optical-thin assumption from \S\ref{column_para}
is valid also for C$^{34}$S we ran LVG radiative transfer models
(RADEX, \citealt{vandertak2007}). We started with the H$_2$ column
densities from Table \ref{source_parameters} and assumed a typical
CS/H$_2$ abundance of $10^{-8}$ with a terrestrial CS/C$^{34}$S ratio
of 23 \citep{wannier1980}. Above the critical density of $2\times
10^7$\,cm$^{-3}$, with the given broad C$^{34}$S spectral FWHM
(between 5 and 12\,km\,s$^{-1}$ for the four sources), the
C$^{34}$(7--6) emission is indeed optically thin.  Hence optical depth
effects are not causing these large offsets.  Since furthermore the
line intensities depend not just on the excitation but more strongly
on the gas column densities, excitation effects only, as quantified in
\S\ref{excitation_effects}, cannot cause the observational offsets as
well.  Therefore, chemical evolution may be more important. A likely
scenario is based on different desorption temperatures of molecules
from dust grains (e.g., \citealt{viti2004}): CS and C$^{34}$S are
desorbed from grains at temperatures of a few 10\,K, and at such
temperatures, these molecules are expected to be well correlated with
the dust continuum emission.  Warming up further, at 100\,K H$_2$O
desorbs and then dissociates to OH. The OH quickly reacts with the
sulphur forming SO and SO$_2$ which then will be centrally peaked (see
\S\ref{modeling}).  Toward G29.96, IRAS\,23151+5912 and
IRAS\,05358+3543 we find that the SO$_2$ emission is centrally peaked
toward the main submm continuum peaks confirming the above outlined
chemical scenario (Fig.~\ref{so2_sample} \&
\S\ref{modeling}\footnote{For G29.96 the SO$_2$ peaks are actually
  right between the better resolved submm continuum sources. This is
  due to the lower resolution of the line data because smoothing the
  continuum to the same spatial resolution, they peak very close to
  the SO$_2$ emission peaks (see, e.g., Fig.~2 in
  \citealt{beuther2007d}).}) .  To further investigate the
C$^{34}$S/SO$_2$ differences, we produced column density ratio maps
between C$^{34}$S and SO$_2$ for G29.96, IRAS\,23151+5912 and
IRAS\,05358+3543, assuming local thermodynamic equilibrium and
optically thin emission (Fig.~\ref{ratio}). To account for the spatial
differences that SO$_2$ is observed toward the submm peak positions
whereas C$^{34}$S is seen more to the edges of the cores, for the
column density calculations we assumed the temperatures given in Table
\ref{source_parameters} for SO$_2$ whereas we used half that
temperature for C$^{34}$S. Although the absolute ratio values are
highly uncertain because of the different spatial filtering properties
of the two molecules, qualitatively as expected, the column density
ratio maps have the lowest values in the vicinity of the submm
continuum sources and show increased emission at the core edges. The
case is less clear for Orion-KL which shows the strongest SO$_2$
emission toward the south-eastern region called compact ridge. Since
this compact ridge is believed to be caused by the interaction of a
molecular outflow with the ambient dense gas (e.g., \citealt{liu2002})
and SO$_2$ is known to be enriched by shock interactions with
outflows, this shock-outflow interaction may dominate in Orion-KL
compared with the above discussed C$^{34}$S/SO$_2$ scenario. For more
details on the chemical evolution see the modeling in
\S\ref{modeling}.

\subsubsection{Nitrogen chemistry}
\label{n}

It is intriguing that we do not detect any nitrogen-bearing molecule
toward the two younger HMPOs IRAS\,23151+5912 and IRAS\,05358+3543
(Fig.\ref{sample_spectra} and Table \ref{linelistall}). This already
indicates that the nitrogen-chemistry needs warmer gas to initiate or
requires more time to proceed. To get an idea about the more subtle
variations of the nitrogen chemistry, one may compare some specific
line pairs: For example, the HN$^{13}$C/CH$_3$CH$_2$CN line blend
(dominated by HN$^{13}$C) and the SO$_2$ line between 348.3 and
348.4\,GHz are of similar strength in the HMC Orion-KL
(Fig.~\ref{sample_spectra}). The same is approximately true for the
HMC G29.96, although SO$_2$ is relatively speaking a bit weaker there.
The more interesting differences arise if one contrasts with the
younger sources. Toward the $10^5$\,L$_{\odot}$ early-HMPO
IRAS\,23151+5912, we only detect the SO$_2$ line and the
HN$^{13}$C/CH$_3$CH$_2$CN line blend remains a non-detection in this
source. In the lower luminosity early-HMPO IRAS\,05358+3543, both
lines are not detected, although another SO$_2$ line at 338.3\,GHz is
detected there.

Judging from these line ratios, one can infer that SO$_2$ is
relatively easy to excite early-on in the evolution of high-mass
star-forming regions. Other sulphur-bearing molecules like H$_2$CS or
CS are released even earlier from the grains, but SO-type molecules
are formed quickly (e.g., \citealt{charnley1997,nomura2004}).  In
contrast to this, the non-detection of spectral lines like the
HN$^{13}$C/CH$_3$CH$_2$CN line blend in the early-HMPOs indicates that
the formation and excitation of such nitrogen-bearing species takes
place in an evolutionary more evolved phase. This may either be due to
molecule-selective temperature-dependent gas-dust desorption processes
or chemical network reactions requiring higher temperatures.
Furthermore, simulations of chemical networks show that the complex
nitrogen chemistry simply needs more time to be activated (e.g.,
\citealt{charnley2001,nomura2004}). Recent modeling by
\citet{garrod2006} indicates that the gradual switch-on phase of hot
molecular cores is an important evolutionary stage to produce complex
molecules. In this picture, the HMCs have switched on their heating
sources earlier and hence had more time to form these nitrogen
molecules.

Comparing just the two HMCs, we find that the CH$_3$CH$_2$CN at
348.55\,GHz is strong in Orion-KL but not detected in G29.96. Since
G29.96 also exhibits less vibrationally-torsionally excited CH$_3$OH
lines, it is likely on average still at lower temperatures than
Orion-KL and may hence have not formed yet all the complex
nitrogen-bearing molecules present already in Orion-KL.

\subsubsection{Modeling}
\label{modeling}

To examine the chemical evolution of warm regions in greater detail we
used the chemical model of \citet{bergin1997}.  This model includes
gas-phase chemistry and the freeze-out to and sublimation from grain
surfaces.   The details of this model are provided in that paper with
the only addition being the inclusion of water ice formation on grain
surfaces.  The binding energies we have adopted are for bare silicate
grains, except for water ice which is assumed to have a binding energy
appropriate for hydrogen bonding between frozen water molecules
\citep{fraser2001}. To explore the chemistry of these hot evaporative
regions we have run the model for 10$^6$\,yrs with starting conditions
at $n_{\rm H_2} = 10^6$\,cm$^{-3}$ and
$T_{\rm{gas}}=T_{\rm{dust}}$=20\,K.  Under these conditions most
gaseous molecules, excluding H$_2$, will freeze onto the grain surface
and the ice mantle forms, dominated by H$_2$O.     This timescale
(10$^5$\,yrs ) is quite short, but is longer than the free-fall time
at this density and is chosen as a representative time that gas might
spend at very high density.  After completion we assume that a massive
star forms and the gas and dust temperature is raised such that the
ice mantle evaporates and the gas-phase chemistry readjusts. We have
made one further adjustment to this model.  Our data suggest that HNC
(a representative nitrogen-bearing species) is not detected in
early-HMPOs and that the release of this species (or its pre-cursor)
from the ice occurs during more evolved and warmer stages. Laboratory
data  on ice desorption suggest that the process is not as simple as
generally used in models where a given molecule evaporates at its
sublimation temperature \citep{collings2004}.  Rather some species
co-desorb with water ice and the key nitrogen-bearing species, NH$_3$,
falls into this category (see also \citealt{viti2004}).  We have
therefore assumed that the ammonia evaporates at the same temperature
as water ice.   Our initial abundances are taken from
\citet{aikawa1996}, except we assume that 50\% of the nitrogen is
frozen in the form of NH$_3$ ice. This assumption is consistent with
two sets of observations.  First, detections of NH$_3$ in ices towards
YSO's find abundances of $\sim$ 2-7\% relative to H$_2$O
\citep{bottinelli2007}.  \citet{dartois2002} find a limit of $\sim$
5\% relative to H$_2$O towards other YSO's.  Our assumed abundance of
NH$_3$ ice is 5 $\times 10^{-6}$ relative to H$_2$O, assuming an ice
abundance of 10$^{-4}$ (as appropriate for cold sources) this provides
an ammonia ice abundance of $\sim$5\%, which is consistent with ice
observations.  Second, high resolution ammonia observations often find
high abundances of NH$_3$ in the gas phase towards hot cores, often as
high as 10$^{-5}$, relative to H$_2$ (e.g., \citealt{cesaroni1994}).
Pure gas phase chemistry will have some difficulty making this high
abundance in the cold phase and we therefore assume it is made on
grains during cold phases.

\begin{figure*}[htb]
  \includegraphics[width=\textwidth]{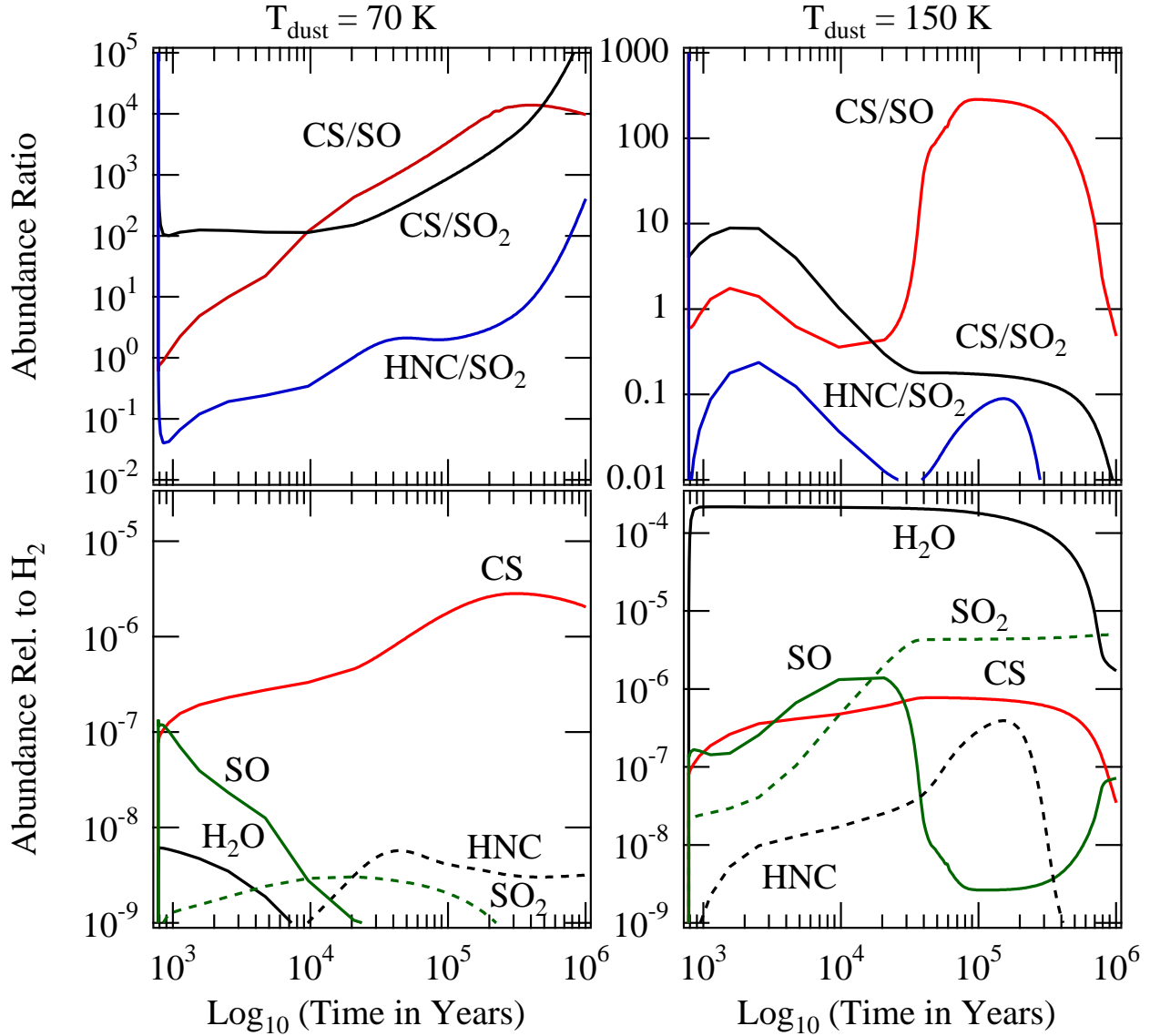}
  \caption{The left and right columns show results from our modeling
    of the ``cold and warm models'', respectively. The two top-panels
    show the abundance ratios between several species important for
    our observations versus the time. The bottom panels present the
    abundances relative to H$_2$. Time in this figure refers to time
    after the onset of massive star formation and hence we do not show
    any evolution during the cold starless phase.}
\label{model}
\end{figure*}

Figure \ref{model} presents our chemical model results at the point of
star ``turn on'' where the gas and dust becomes warmer and ices
evaporate.     Two different models were explored.  The first labeled
as $T_{dust}$=70\,K (``warm model'') is a model which is insufficient
to evaporate water ice, while the second, ``hot model''
with $T_{dust}$=150\,K, evaporates the entire ice mantle.   Much of
the chemical variations amongst the early-HMPO and HMC phases is found
for CS, SO, SO$_2$, and HNC (note this line is blended with
CH$_3$CH$_2$CN) and we focus on these species in our plots (along with
H$_2$O). It is also important to reiterate that due to differences in
spatial sampling between the different sources as well as between line
and continuum emission, we cannot derive accurate abundances from
these data, but rather can attempt to use the models to explain
trends. For the warm model the main result is that the imbalance
created in the chemistry by the release of most species, with ammonia
and water remaining as ices on the grains, leads to enhanced
production of CS.   Essentially the removal of oxygen from the gas
(excluding the O found in CO) allows for rapid CS production from the
atomic Sulfur frozen on grains during the cold phase. Thus, for
early-HMPOs, which might not have a large fraction of gas above the
water sublimation point ($T\sim110$\,K), the ratios of CS to other
species are quite large. In the hot model, when the temperature can
evaporate both H$_2$O and NH$_3$ ice, ratios between the same
molecules are orders of magnitude lower (Fig.\,\ref{model}). There is
a large jump in the water vapor abundance (and NH$_3$, which is not
shown) between the warm and the hot model driving the chemistry into
new directions.  CS remains in the gas, but not with as high abundance
as in the warm phase and is gradually eroded into SO and ultimately
SO$_2$. Hence, SO$_2$ appears to be a better tracer of more evolved
stages. In this sense, even the early-HMPOs can be considered as
relatively evolved, and it will be important to extend similar studies
to even earlier evolutionary stages.  HNC also appears as a brief
intermediate product of the nitrogen chemistry.

The above picture is in qualitative agreement with our observations:
that CS should be a good tracer in early evolutionary states even
prior to our observed sample, but that it is less well suited for more
evolved regions like those studied here.  Other sulfur-bearing
molecules, in particular SO$_2$, appear to better trace the warmest
gas near the forming star.   HNC, and perhaps other nitrogen-bearing
species, are better tracers when the gas is warm enough to evaporate a
significant amount of the ice mantle.

\subsection{Searching for disk signatures}
\label{disks}

While the chemical evolution of massive star-forming regions is
interesting in itself, one also wants to use the different
characteristics of molecular lines as tools to trace various physical
processes. While molecules like CO or SiO have regularly been used to
investigate molecular outflows (e.g., \citealt{arce2006}), the problem
to identify the right molecular tracer to investigate accretion disks
in massive star formation is much more severe (e.g.,
\citealt{cesaroni2006}). Major observational obstacles arise from the
fact that disk-tracing molecular lines are usually often not
unambiguously found only in the accretion disk, but that other
processes can produce such line emission as well. For example,
molecular lines from CN and HCN are high-density tracers and were
believed to be good candidates to investigate disks in embedded
protostellar sources (e.g., \citealt{aikawa2001}). However,
observations revealed that both spectral lines are strongly affected
by the associated molecular outflow and hence difficult to use for
massive disk studies (CN \citealt{beuther2004e}, HCN
\citealt{zhang2007}).

\begin{figure*}[htb]
\includegraphics[angle=-90,width=6.5cm]{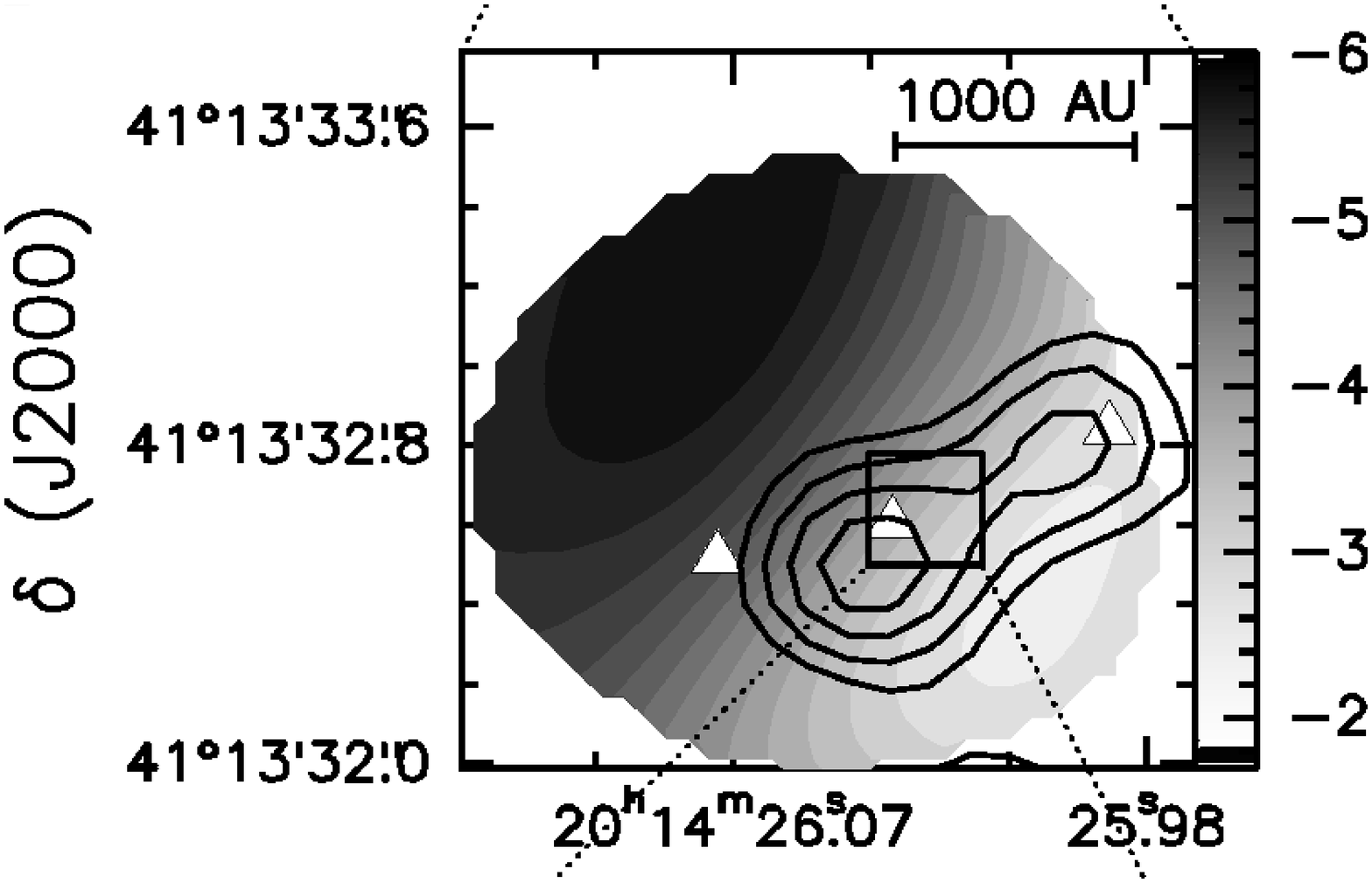}
\includegraphics[angle=-90,width=4.5cm]{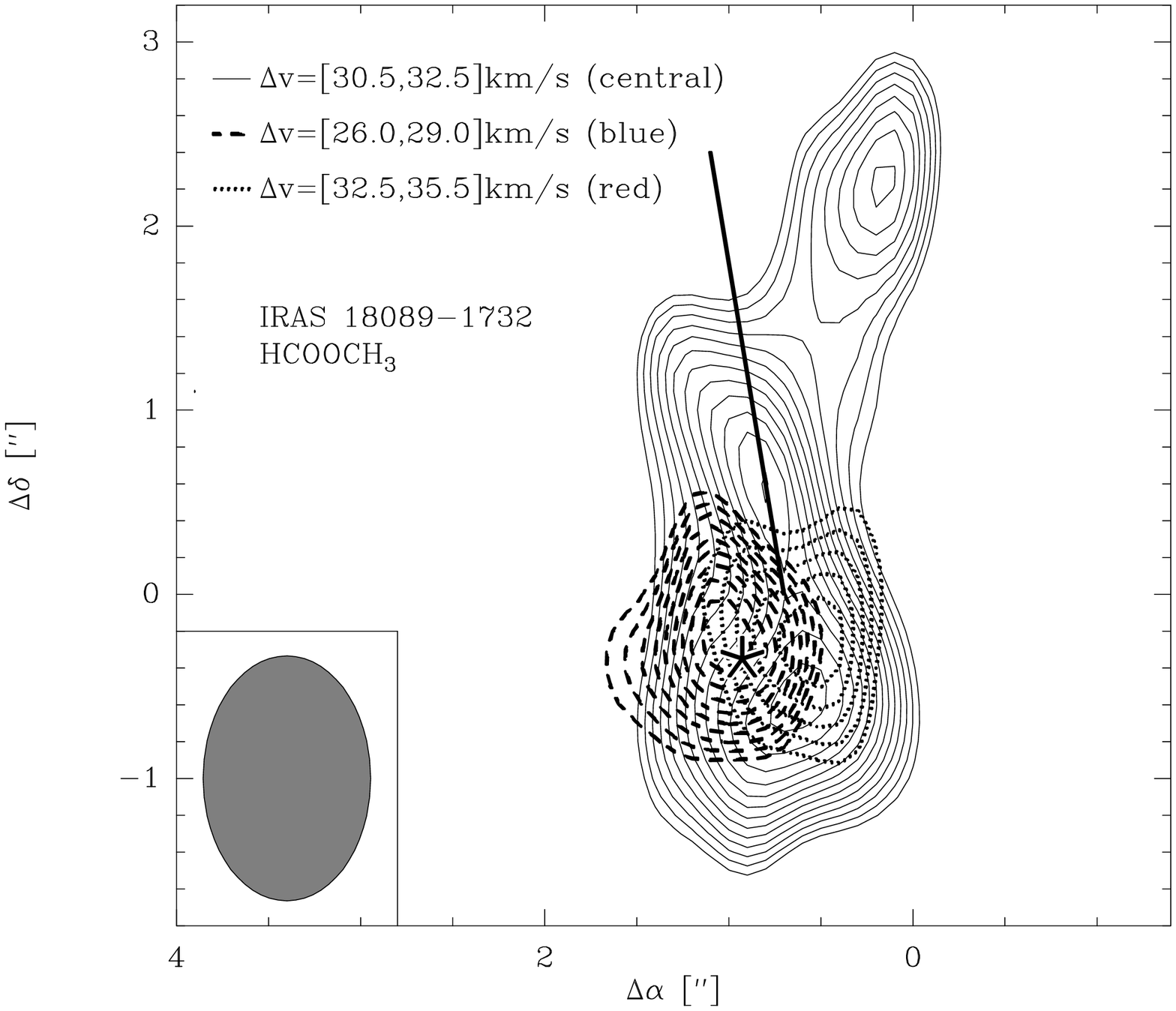}
\includegraphics[angle=-90,width=5.0cm]{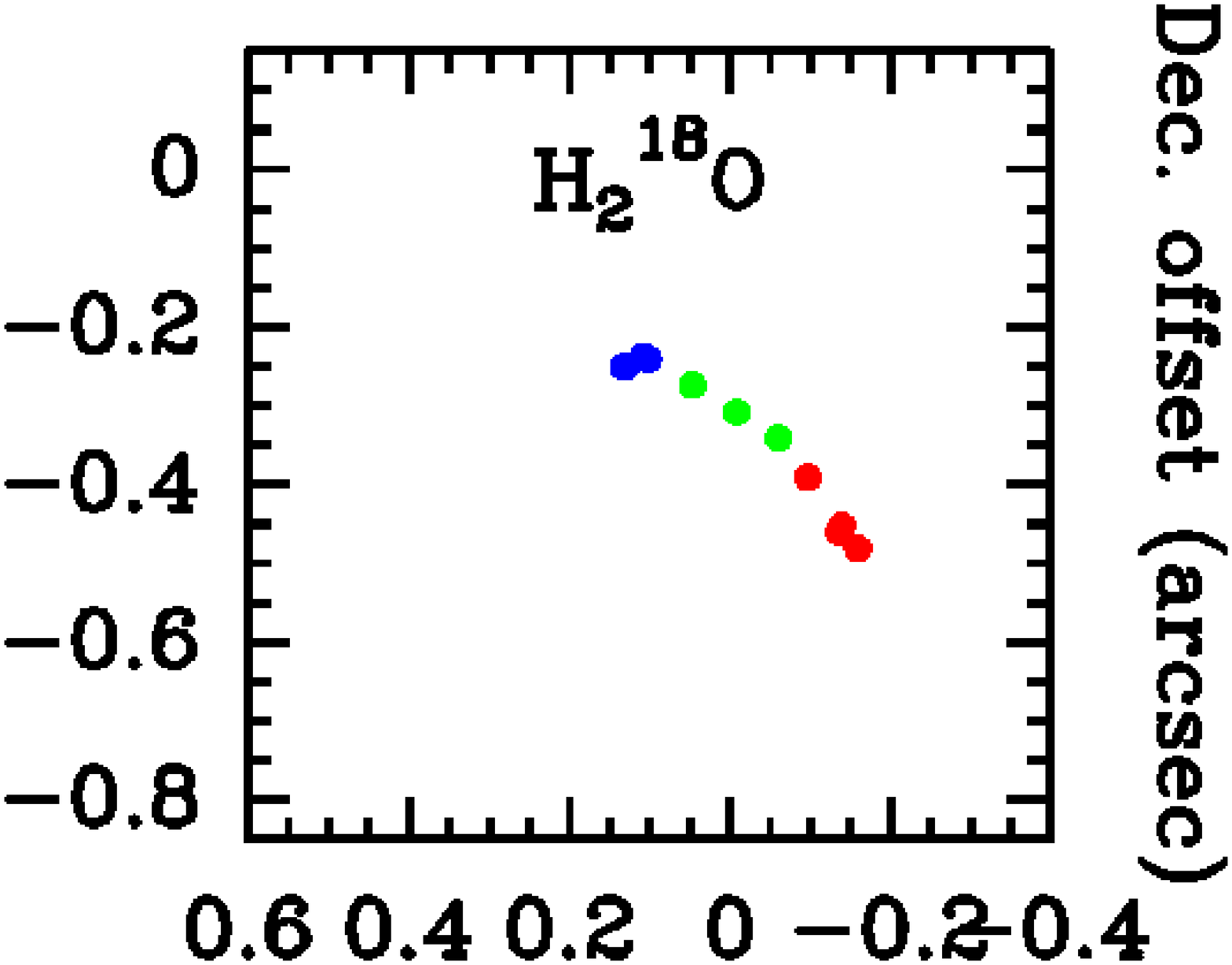}\\
\hspace*{4.0cm}\includegraphics[angle=-90,width=4.5cm]{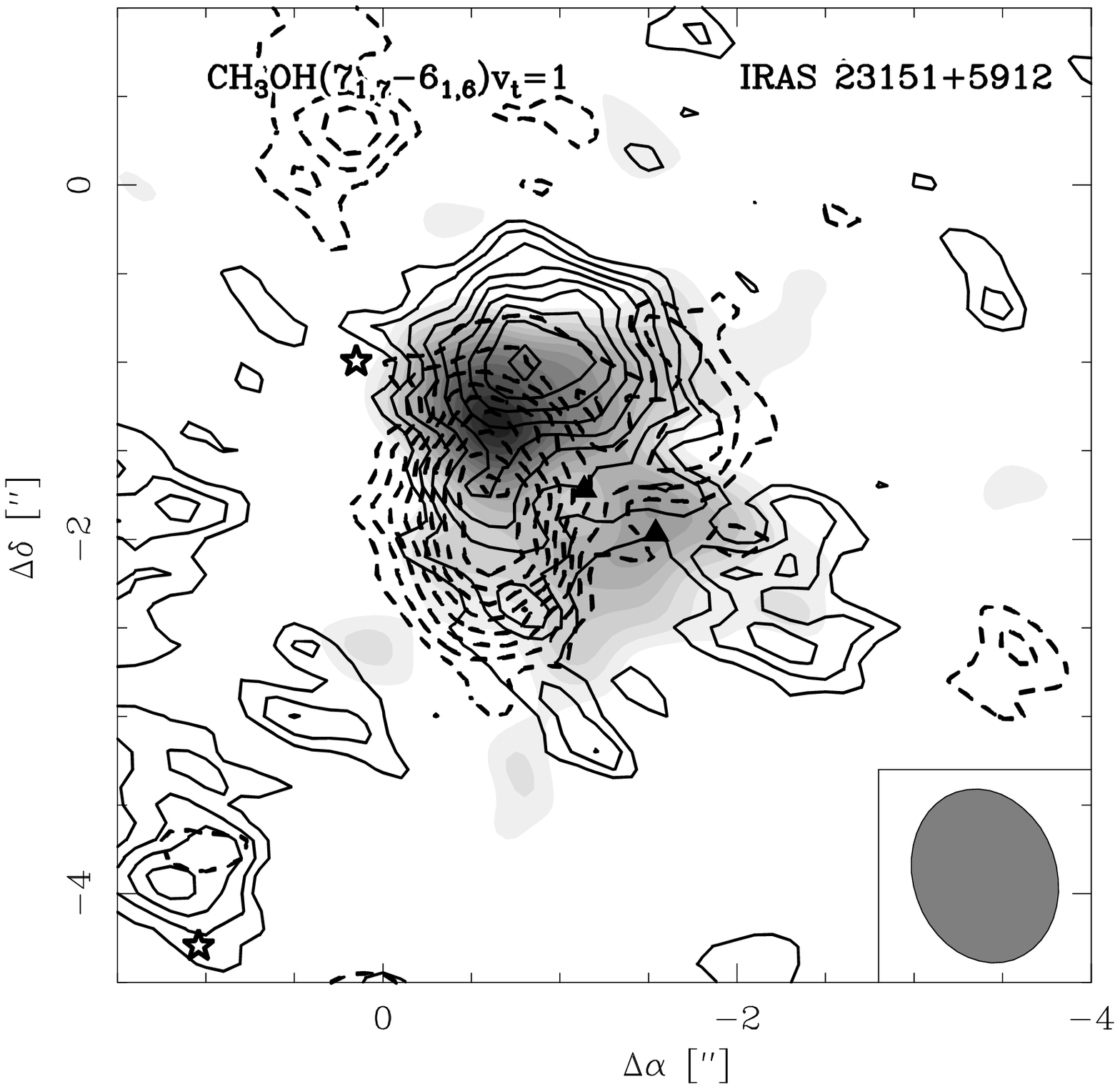}
\includegraphics[angle=-90,width=5.2cm]{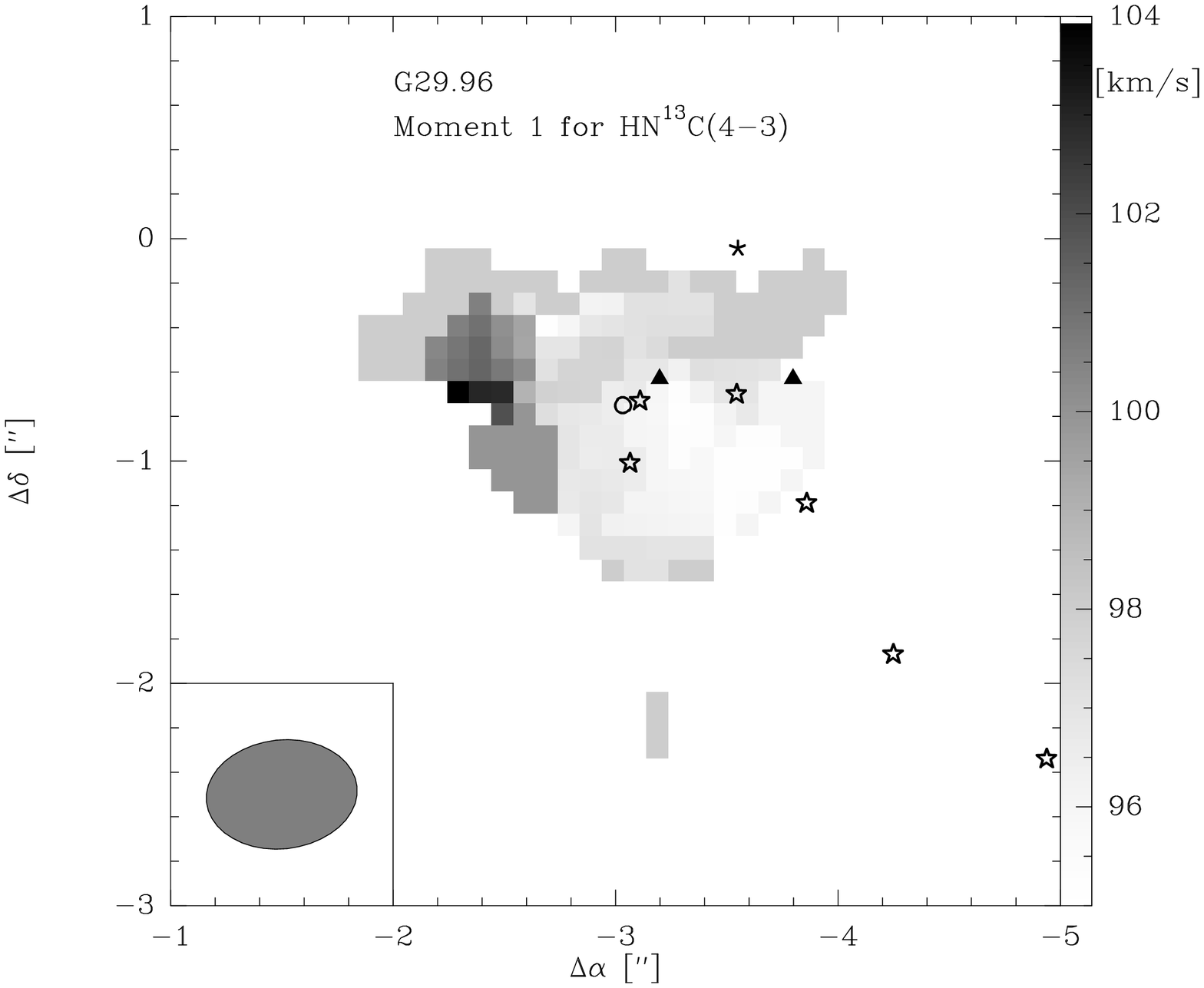}
\caption{Examples of rotation-tracing molecules: top-left: C$^{34}$S
  in IRAS\,20126+4104 \citep{cesaroni1999,cesaroni2005}, top-middle:
  HCOOCH$_3$ in IRAS\,18089-1732 \citep{beuther2005c}, top-right:
  H$_2^{18}$O in AFGL2591 \citep{vandertak2006}, bottom-left: CH$_3$OH
  $v_t=1$ in IRAS\,23151+5912 \citep{beuther2007f}, bottom-right:
  HN$^{13}$C in G29.96 \citep{beuther2007d}.}
\label{disk_examples}
\end{figure*}

As presented in \citet{cesaroni2006}, various different molecules have
in the past been used to investigate disks/rotational signatures in
massive star formation (e.g., CH$_3$CN, C$^{34}$S, NH$_3$, HCOOCH$_3$,
C$^{17}$O, H$_2^{18}$O, see also Fig.~\ref{disk_examples}). The data
presented here add three other potential disk tracers (HN$^{13}$C and
HC$_3$N for G29.96, \citealt{beuther2007d}, and torsionally excited
CH$_3$OH for IRAS\,23151+5912 and IRAS\,05358+3543
\citealt{beuther2007f,leurini2007}, Fig.~\ref{disk_examples}). An
important point to note is that in most sources only one or the other
spectral line exclusively allows a study of rotational motions,
whereas other lines apparently do not trace the warm disks. For
example, C$^{34}$S traces the Keplerian motion in the young source
IRAS\,20126+4104 whereas it does not trace the central protostars at
all in the sources presented here (Fig.~\ref{c34s_sample}).  In the
contrary, HN$^{13}$C shows rotational signatures in the HMC G29.96,
but it remains completely undetected in the younger early-HMPO sources
of our sample (Figs.~\ref{disk_examples} \& \ref{sample_spectra}). As
discussed in sections \ref{c34s} and \ref{n}, this implies that,
depending on the chemical evolution, molecules like C$^{34}$S should
be better suited for disk studies at very early evolutionary stages,
whereas complex nitrogen-bearing molecules are promising in more
evolved hot-core-type sources. While the chemical evolution is
important for these molecules, temperature effects have to be taken
into account as well.  For example, the torsionally excited CH$_3$OH
line traces rotating motions in IRAS\,23151+5912 and IRAS\,05358+3543
(\citealt{beuther2007f,leurini2007}, Fig.~\ref{disk_examples}) but it
is weak and difficult to detect in colder and younger sources.
Therefore, some initial heating is required to employ highly excited
lines for kinematic studies.  In addition to these evolutionary
effects, optical depth is important for many lines. For example,
\citet{cesaroni1997,cesaroni1999} have shown that CH$_3$CN traces the
rotating structure in IRAS\,20126+4104, whereas the same molecule does
not indicate any rotation in IRAS\,18089-1732 \citep{beuther2005c}. A
likely explanation for the latter is high optical depth of the
CH$_3$CN submm lines \citep{beuther2005c}.  Again other molecules are
excited in the accretion disks as well as the surrounding envelope,
causing confusion problems to disentangle the various physical
components. {\it In summary, getting a chemical rich spectral line
  census like the ones presented here shows that one can find several
  disk-tracing molecules in different sources, but it also implies
  that some previously assumed good tracers are not necessarily
  universally useful.}

The advent of broad bandpass interferometers like the SMA now
fortunately allows to observe many molecular lines simultaneously.
This way, one often finds a suitable rotation-tracing molecule in an
observational spectral setup. Nevertheless, one has to keep in mind
the chemical and physical complexity in such regions, and it is likely
that in many cases only combined modeling of infall, rotation and
outflow will disentangle the accretion disk from the rest of the
star-forming gas and dust core.

\section{Conclusion and Summary}

We compiled a sample of four massive star-forming regions in different
evolutionary stages with varying luminosities that were observed in
exactly the same spectral setup at high angular resolution with the
SMA. We estimated column densities for all sources and detected
species, and we compared the spatial distributions of the molecular
gas.  This allows us to start investigating chemical evolutionary
effects also in a spatially resolved manner. Chemical modeling was
conducted to explain our observations in more detail.

A general result from this comparison is that many different physical
and chemical processes are important to produce the complex chemical
signatures we observe. While some features, e.g., the non-detection of
the rich vibrationally-torsionally excited CH$_3$OH line forest toward
the two early-HMPOs can be explained by on average lower temperatures of
the molecular gas compared to the more evolved HMCs, other
observational characteristics require chemical evolutionary sequences
caused by heating including grain-surface and gas phase reactions.
Even other features are then better explained by shock-induced
chemical networks.

The rare isotopologue C$^{34}$S is usually not detected right toward
the main submm continuum peaks, but rather at the edge of the
star-forming cores. This may be explained by temperature-selective
gas-desorption processes and successive gas chemistry networks.
Furthermore, we find some nitrogen-bearing molecular lines to be only
present in the HMCs, whereas they remain undetected at earlier
evolutionary stages. This indicates that the formation and excitation
of many nitrogen-bearing molecules needs considerably higher
temperatures and/or more time during the warm-up phase of the HMC,
perhaps relating to the fact that NH$_3$ is bonded within the water
ice mantle.  Although the statistical database is still too poor to
set tighter constraints, these observations give the direction how one
can use the presence {\it and} morphology of various molecular lines
to identify and study different (chemical) evolutionary sequences.

Furthermore, we discussed the observational difficulty to
unambiguously use one or the other spectral line as a tracer of
massive accretion disks. While some early spectral line candidates are
discarded for such kind of studies by now (e.g., CN), in many other
sources we find different lines exhibiting rotational velocity
signatures. The observational feature that in most sources apparently
only one or the other spectral line exclusively traces the desired
structures has likely to be attributed to a range of effects. (1)
Chemical effects, where for example C$^{34}$S may work in the youngest
sources whereas some nitrogen-bearing molecules like HN$^{13}$C are
better in typical HMCs. (2) Confusion from multiple gas components,
mainly outflows, infall from the envelope and rotation.  (3) High
optical depth from many molecular lines. This implies that for future
statistical studies we have to select spectral setups that comprise
many molecular lines from various species. This way, one has good
chances to identify for each source separately the right molecular
tracer, and hence still draw statistically significant conclusions.

To advance in this field and to become more quantitative, different
steps are necessary. First of all, we need to establish a larger
database of more sources at different evolutionary stages, in
particular even younger sources, as well as with varying luminosities
to better characterize the differences and similarities.  From an
observational and technical point of view, although the presented data
are state of the art multi-wavelength and high angular resolution
observations, the quantitative interpretation is still hampered by the
spatial filtering of the interferometer. To become more quantitative,
it is therefore necessary to complement such data with the missing
short spacing information. While we have high angular resolution in
all datasets with a similar baseline coverage and hence similarly
covered angular scales, the broad range of distances causes a
different coverage of sampled linear spatial scales.  Hence the
missing short spacings affect each dataset in a different fashion
which is currently the main limiting factor for a better quantitative
interpretation of the data.  Therefore, obtaining single-dish
observations in the same spectral setup and then combining them with
the SMA observations is a crucial step to derive more reliable column
densities and from that abundances. These parameters then can be used
by theorists to better model the chemical networks, explain the
observations and predict other suitable molecules for, e.g., massive
disk studies.

\begin{acknowledgements} 
  H.B.~acknowledges financial support by the Emmy-Noether-Program of the
  Deutsche Forschungsgemeinschaft (DFG, grant BE2578).
\end{acknowledgements}

%\bibliography{/home/beuther/tex/bibliography}   
%\bibliography{/Users/henrikbeuther/paper/bibliography}
%\bibliographystyle{aa}    % this does the style, aa.bst necessary

\begin{longtable}[htb]{lrrcccc}
  \caption{Line peak intensities $S_{\rm{peak}}$ and upper state
    energy levels $E_u/k$ from spectra toward peak positions of the
    respective massive star-forming regions (Fig.~\ref{sample_spectra}).}\\
  \hline \hline
  Freq. & Line & $E_u/k$ & $S_{\rm{peak}}$ & $S_{\rm{peak}}$ & $S_{\rm{peak}}$  & $S_{\rm{peak}}$  \\
  (GHz) &      & (K)     & (Jy)            & (Jy)            & (Jy)             & (Jy)\\
  &      &         & Orion & G29.96 & 23151 & 05358 \\
  \hline
  337.252 & CH$_3$OH$(7_{3,5}-6_{3,4})$A($v_t$=2)        & 739 & 6.0 \\
  337.274 & CH$_3$OH$(7_{4,3}-6_{4,2})$A($v_t$=2)        & 695 & 7.4 \\
  337.279 & CH$_3$OH$(7_{2,5}-6_{2,4})$E($v_t$=2)        & 727 & 5.4 \\
  337.284 & CH$_3$OH$(7_{0,7}-6_{0,6})$A($v_t$=2)        & 589 & 9.9 \\
  337.297 & CH$_3$OH$(7_{1,7}-6_{1,6})$A($v_t$=1)        & 390 & 10.6 & 1.7 \\
  337.312 & CH$_3$OH$(7_{1,6}-6_{1,5})$E($v_t$=2)        & 613 & 9.2 \\
  337.348 & CH$_3$CH$_2$CN$(38_{3,36}-37_{3,35})$        & 328 & 14.2 & 1.5 & & \\
  337.397 & C$^{34}$S(7--6)                              & 65  & 12.6 & 2.0 & 0.3 & 0.6 \\
  337.421 & CH$_3$OCH$_3(21_{2,19}-20_{3,18})$           & 220 & 3.0 & 0.6 \\
  337.446 & CH$_3$CH$_2$CN$(37_{4,33}-36_{4,32})$        & 322 & 11.2 & 0.8 \\
  337.464 & CH$_3$OH$(7_{6,1}-6_{0,0})$A($v_t$=1)        & 533 & 7.2 & 1.1 \\
  337.474 & UL                                           &     & 4.9 & 0.6 \\
  337.490 & HCOOCH$_3(27_{8,20}-26_{8,19})$E             & 267 & 6.3 & 0.7 \\
  337.519 & CH$_3$OH$(7_{5,2}-6_{5,2})$E($v_t$=1)        & 482 & 8.1 & 1.0 \\
  337.546 & CH$_3$OH$(7_{5,3}-6_{5,2})$A($v_t$=1)        & 485 & 10.0$^b$ & 1.4$^b$ & & \\
  & CH$_3$OH$(7_{5,2}-6_{5,1})$A$^{-}$($v_t$=1)          & 485 & 10.0$^b$ & 1.4$^b$ & & \\
  337.582 & $^{34}$SO$(8_8-7_7)$                         & 86  & 12.2 & 2.0 & 1.1 &\\
  337.605 & CH$_3$OH$(7_{2,5}-6_{2,4})$E($v_t$=1)        & 429 & 9.7 & 2.4 & & \\
  337.611 & CH$_3$OH$(7_{6,1}-6_{6,0})$E($v_t$=1)        & 657 & 6.2$^b$ & 2.0$^b$ & & \\
  & CH$_3$OH$(7_{3,4}-6_{3,3})$E($v_t$=1)                & 388 & 6.2$^b$ & 2.0$^b$ & & \\
  337.626 & CH$_3$OH$(7_{2,5}-6_{2,4})$A($v_t$=1)        & 364 & 11.0 & 1.9 & & \\
  337.636 & CH$_3$OH$(7_{2,6}-6_{2,5})$A$^-$($v_t$=1)    & 364 & 8.2 & 2.5 & & \\
  337.642 & CH$_3$OH$(7_{1,7}-6_{1,6})$E($v_t$=1)        & 356 & 10.9$^b$ & 2.9$^b$ & 0.6$^b$ & 1.1$^b$\\
  337.644 & CH$_3$OH$(7_{0,7}-6_{0,6})$E($v_t$=1)        & 365 & 10.9$^b$ & 2.9$^b$ & 0.6$^b$ & 1.1$^b$\\
  337.646 & CH$_3$OH$(7_{4,3}-6_{4,2})$E($v_t$=1)        & 470 & 10.9$^b$ & 2.9$^b$ & 0.6$^b$ & 1.1$^b$ \\
  337.648 & CH$_3$OH$(7_{5,3}-6_{5,2})$E($v_t$=1)        & 611 & 10.9$^b$ & 2.9$^b$ & 0.6$^b$ & 1.1$^b$ \\
  337.655 & CH$_3$OH$(7_{3,5}-6_{3,4})$A($v_t$=1)        & 461 & 10.8$^b$ & 2.0$^b$ & & \\
  & CH$_3$OH$(7_{3,4}-6_{3,3})$A$^-$($v_t$=1)            & 461 & 10.8$^b$ & 2.0$^b$ & & \\
  337.671 & CH$_3$OH$(7_{2,6}-6_{2,5})$E($v_t$=1)        & 465 & 10.2 & 2.1 & & \\
  337.686 & CH$_3$OH$(7_{4,3}-6_{4,2})$A($v_t$=1)        & 546 & 9.$^b$5 & 2.0$^b$ & & \\
  & CH$_3$OH$(7_{4,4}-6_{4,3})$A$^-$($v_t$=1)            & 546 & 9.5$^b$ & 2.0$^b$ & & \\
  & CH$_3$OH$(7_{5,2}-6_{5,1})$E($v_t$=1)                & 494 & 9.5$^b$ & 2.0$^b$ & & \\
  337.708 & CH$_3$OH$(7_{1,6}-6_{1,5})$E($v_t$=1)        & 489 & 7.9 & 1.8 & & \\
  337.722 & CH$_3$OCH$_3(7_{4,4}-6_{3,3})$EE             & 48  & & 0.9 & & \\
  337.732 & CH$_3$OCH$_3(7_{4,3}-6_{3,3})$EE             & 48  & & 1.4 & &\\
  337.749 & CH$_3$OH$(7_{0,7}-6_{0,6})$A($v_t$=1)        & 489 & 8.7 & 1.9 & & \\
  337.778 & CH$_3$OCH$_3(7_{4,4}-6_{3,4})$EE             & 48  & & 1.3 & &\\
  337.787 & CH$_3$OCH$_3(7_{4,3}-6_{3,4})$AA             & 48  & & 1.4 & &\\
  337.825 & HC$_3$N$(37-36)v_7=1$                        & 629 & 14.8 & 1.4 & & \\
  337.838 & CH$_3$OH$(20_{6,14}-21_{5,16})$E             & 676 & 5.6 & 1.1 & & \\
   337.878 & CH$_3$OH$(7_{1,6}-6_{1,5})$A($v_t$=2)        & 748 & 2.7 & 0.6 & & \\
  337.969 & CH$_3$OH$(7_{1,6}-6_{1,5})$A($v_t$=1)        & 390 & 12.0 & 2.1 & & \\
  338.081 & H$_2$CS$(10_{1,10}-9_{1,9})$                 & 102 & 5.8 & 2.3 & & 0.5 \\
  338.125 & CH$_3$OH$(7_{0,7}-6_{0,6})$E                 & 78  & 6.9 & 2.8 & 1.4 & 1.9 \\
  338.143 & CH$_3$CH$_2$CN$(37_{3,34}-36_{3,33})$        & 317 & 14.4 & 0.9 & & \\
  338.214 & CH$_2$CHCN$(37_{1,37}-36_{1,36})$            & 312 & 4.0 \\
  338.306 & SO$_2(18_{4,1}-18_{3,1})$                  & 197 & x$^c$ & 0.8 & 1.2 & 0.7 \\
  338.345 & CH$_3$OH$(7_{1,7}-6_{1,6})$E                 & 71  & 13.4 & 2.1 & 1.3 & 2.3\\
  338.405 & CH$_3$OH$(7_{6,2}-6_{6,1})$E                 & 244 & 13.1$^b$ & 3.0$^b$ &  &  \\
  338.409 & CH$_3$OH$(7_{0,7}-6_{0,6})$A                 & 65  & 13.1$^b$ & 3.0$^b$ & 1.5 & 2.4\\
  338.431 & CH$_3$OH$(7_{6,1}-6_{6,0})$E                 & 254 & 9.5 & 1.8 &  &  \\
  338.442 & CH$_3$OH$(7_{6,1}-6_{6,0})$A                 & 259 & 11.7$^b$ & 2.7$^b$ &  & 0.5$^b$ \\
  & CH$_3$OH$(7_{6,2}-6_{6,1})$A$^-$                     & 259 & 11.7$^b$ & 2.7$^b$ &  & 0.5$^b$ \\
  338.457 & CH$_3$OH$(7_{5,2}-6_{5,1})$E                 & 189 & 8.9 & 2.0 & 0.4 & 0.6 \\
  338.475 & CH$_3$OH$(7_{5,3}-6_{5,2})$E                 & 201 & 12.6 & 2.5 & 0.5 & \\
  338.486 & CH$_3$OH$(7_{5,3}-6_{5,2})$A                 & 203 & 9.4$^b$ & 2.3$^b$ & 0.8$^b$ & 0.8$^b$ \\
  & CH$_3$OH$(7_{5,2}-6_{5,1})$A$^-$                     & 203 & 9.4$^b$ & 2.3$^b$ & 0.8$^b$ & 0.8$^b$ \\
  338.504 & CH$_3$OH$(7_{4,4}-6_{4,3})$E                 & 153 & 8.5 & 2.7 & 0.7 & 0.7 \\
  338.513 & CH$_3$OH$(7_{4,4}-6_{4,3})$A$^-$             & 145 & 13.7$^b$ & 2.8$^b$ & 1.4$^b$ & 1.3$^b$ \\
  & CH$_3$OH$(7_{4,3}-6_{4,2})$A                         & 145 & 13.7$^b$ & 2.8$^b$ & 1.4$^b$ & 1.3$^b$ \\
  & CH$_3$OH$(7_{2,6}-6_{2,5})$A$^-$                     & 103 & 13.7$^b$ & 2.8$^b$ & 1.4$^b$ & 1.3$^b$ \\
  338.530 & CH$_3$OH$(7_{4,3}-6_{4,2})$E                 & 161 & 5.8 & 2.7 & 0.7 & 0.9 \\
  338.541 & CH$_3$OH$(7_{3,5}-6_{3,4})$A$^+$             & 115 & 12.5$^b$ & 3.0$^b$ & 2.0$^b$ & 1.4$^b$ \\
  338.543 & CH$_3$OH$(7_{3,4}-6_{3,3})$A$^-$             & 115 & 12.5$^b$ & 3.0$^b$ & 2.0$^b$ & 1.4$^b$ \\
  338.560 & CH$_3$OH$(7_{3,5}-6_{3,4})$E                 & 128 & 15.6 & 2.5 & 0.9 & 0.6 \\
  338.583 & CH$_3$OH$(7_{3,4}-6_{3,3})$E                 & 113 & 11.5 & 3.4 & 1.0 & 1.1 \\
  338.612 & SO$_2(20_{1,19}-19_{2,18})$                  & 199 & x$^c$ & 2.9 & 1.5 & 1.9 \\
  338.615 & CH$_3$OH$(7_{1,6}-6_{1,5})$E                 & 86  & x$^d$ & 2.9$^d$ & 1.5$^d$ & 1.9$^d$\\
  338.640 & CH$_3$OH$(7_{2,5}-6_{2,4})$A                 & 103 & 7.2 & 2.5 & 1.0 & 1.0 \\
  338.722 & CH$_3$OH$(7_{2,5}-6_{2,4})$E                 & 87  & 10.2$^b$ & 3.2$^b$ & 2.1$^b$ & 2.7$^b$\\
  338.723 & CH$_3$OH$(7_{2,6}-6_{2,5})$E                 & 91  & 10.2$^b$ & 3.2$^b$ & 2.1$^b$ & 2.7$^b$\\
  338.760 & $^{13}$CH$_3$OH$(13_{7,7}-12_{7,6})$A        & 206 & 4.0 & 1.1 & & \\
  338.769 & HC$_3$N$(37-36)v_7=2$                        & 525 & ? & ? & & \\
  338.786 & $^{34}$SO$_2(14_{4,10}-14_{3,1})$            & 134 & 6.2 \\
  338.886 & C$_2$H$_5$OH$(15_{7,8}-15_{6,19})$           & 162 & 5.3 & 0.8 & & \\
  338.930 & $^{30}$SiO(8--7)                             & 73  & 24.4 \\
  339.058 & C$_2$H$_5$OH$(14_{7,7}-14_{6,8})$            & 150 & x$^e$ & 0.6 & & \\
  347.232 & CH$_2$CHCN$(38_{1,38}-37_{1,37})$            & 329 & 4.9 & 0.6 & & \\
  347.331 &  $^{28}$SiO(8--7)                            & 75  & 22.1 & 0.9 & 0.7 &\\
  347.438 & UL                                           &     & 7.5 & & & \\
  347.446 & UL                                           &     & 3.4 & 0.8 & & \\
  347.478 &  HCOOCH$_3(27_{1,26}-26_{1,25})$E            & 247 & 4.8 \\
  347.494 &  HCOOCH$_3(27_{5,22}-26_{5,21})$A            & 247 & 2.9 & 0.6 & & \\
  347.590 &  HCOOCH$_3(16_{6,10}-15_{5,11})$A            & 104 & 2.4 \\
  347.599 &  HCOOCH$_3(16_{6,10}-15_{5,11})$E            & 105 & 1.7 \\
  347.617 &  HCOOCH$_3(28_{10,19}-27_{10,18})$A          & 307 & 2.6 \\
  347.628 &  HCOOCH$_3(28_{10,19}-27_{10,18})$E          & 307 & 3.6 \\
  347.667 &  UL                                          &     & 4.5 \\
  347.759 &  CH$_2$CHCN$(36_{2,34}-35_{2,32})$           & 317 & 8.3 & 0.7 & & \\
  347.792 & UL                                           &     & 5.4 & 0.7 & & \\
  347.842 & UL, $^{13}$CH$_3$OH                          &     & 3.1 & 0.5 & & \\
  347.916 &  C$_2$H$_5$OH$(20_{4,17}-19_{4,16})$         & 251 & 3.3 & 0.7 & & \\
  347.983 &  UL                                          &     & & 0.6 & & \\
  348.050 &  HCOOCH$_3(28_{4,24}-27_{4,23})$E            & 266 & 2.8 \\
  348.066 &  HCOOCH$_3(28_{6,23}-27_{6,22})$A            & 266 & 3.0 \\
  348.118 &  $^{34}$SO$_2(19_{4,16}-19_{3,17})$          & 213 & 5.8 \\
  348.261 &  CH$_3$CH$_2$CN$(39_{2,37}-38_{2,36})$       & 344 & 11.2 & 1.2 & & \\
  348.340 &  HN$^{13}$C(4--3)                            & 42  & 16.1$^b$ & 2.0$^b$ & &\\
  348.345 &  CH$_3$CH$_2$CN$(40_{2,39}-39_{2,38})$       & 351 & 16.1$^b$ & 2.0$^b$ & & \\
  348.388 & SO$_2(24_{2,22}-23_{3,21})$                  & 293 & 9.3 & 0.5 & 1.0 & \\
  348.518 & UL, HNOS$(1_{1,1}-2_{0,2})$                  &     & 10.6 & 0.7 \\
  348.532 &  H$_2$CS$(10_{1,9}-9_{1,8})$                 & 105 & 7.4 & 1.9 & & \\
  348.553 &  CH$_3$CH$_2$CN$(40_{1,39}-39_{1,38})$       & 351 & 20.1 \\
  348.910 &  HCOOCH$_3(28_{9,20}-27_{9,19})$E            & 295 & 11.0$^b$ & 1.6$^b$ & & \\
  348.911 & CH$_3$CN$(19_{9}-18_{9})$                    & 745 & 11.0$^b$ & 1.6$^b$ & & \\
  348.991 & CH$_2$CHCN$(37_{1,36}-36_{1,35})$            & 325 & 5.9 \\
  349.025 & CH$_3$CN$(19_{8}-18_{8})$                    & 624 & 9.5 & 1.1 \\
  349.107 & CH$_3$OH$(14_{1,13}-14_{0,14})$              & 43  & 12.2 & 3.1 & 1.3 & 1.1\\
  \hline \hline
  \multicolumn{7}{l}{\footnotesize $^a$ Doubtful detection since other close $v_t=2$ lines with similar upper }\\
  \multicolumn{7}{l}{\footnotesize energy levels were not detected.}\\
  \multicolumn{7}{l}{\footnotesize $^b$ Line blend.}\\
  \multicolumn{7}{l}{\footnotesize $^c$ No flux measurement possible because averaged over the given 5700\,AU negative }\\
  \multicolumn{7}{l}{\footnotesize features due to missing short spacings overwhelm the positive features (see Figs.~\ref{ch3oh_sample} \& \ref{so2_sample}).}\\
  \multicolumn{7}{l}{\footnotesize $^d$ Peak flux corrupted by neighboring SO$_2$ line.}\\
  \multicolumn{7}{l}{\footnotesize $^e$ Only detectable with higher
    spatial resolution \citep{beuther2005a}.}\label{linelistall}
\end{longtable}

\begin{table}[htb]
\caption{Molecular column densities.}
\begin{tabular}{lrrrr}
\hline
\hline
                   & Orion-KL$^a$ & G29.96  & 23151 & 05358 \\
\hline
CH$_3$OH           & $2{\sf x}10^{16}$ &  $4{\sf x}10^{17}\,^b$ & $3{\sf x}10^{16}-1{\sf x}10^{17}\,^c$   &  $4{\sf x}10^{15}-4{\sf x}10^{18}\,^d$  \\
CH$_3$CH$_2$CN     & $5{\sf x}10^{15}$ & $1{\sf x}10^{16}$ & --    & -- \\
CH$_2$CHCN         & $5{\sf x}10^{15}$ & $7{\sf x}10^{15}$ & --    & -- \\
C$^{34}$S          & $2{\sf x}10^{14}$ & $2{\sf x}10^{15}$ & $2{\sf x}10^{14}$ & $6{\sf x}10^{13}$ \\
CH$_3$OCH$_3$      & $1{\sf x}10^{16}$ & $2{\sf x}10^{17}\,^e$ & --    & -- \\
HCOOCH$_3$         & $6{\sf x}10^{15}$ & $8{\sf x}10^{16}$ & --    & -- \\
$^{34}$SO          & $6{\sf x}10^{14}$ & $1{\sf x}10^{16}$ & $4{\sf x}10^{15}$ & -- \\
SO$_2$             & $2{\sf x}10^{15}$ & $3{\sf x}10^{16}$ & $3{\sf x}10^{16}$ & $1{\sf x}10^{16}\,^d$ \\
HC$_3$N            & $6{\sf x}10^{14}$ & $2{\sf x}10^{15}$ & --    & -- \\
HN$^{13}$C         & blend    & blend   & --    & -- \\
H$_2$CS            & $4{\sf x}10^{14}$ & $2{\sf x}10^{16}$ & --    & $4{\sf x}10^{14}$  \\
C$_2$H$_5$OH       & $2{\sf x}10^{15}$ & $6{\sf x}10^{16}$ & --    & -- \\
SiO                & $2{\sf x}10^{14}$ & $4{\sf x}10^{14}$ & $2{\sf x}10^{14}$ & -- \\ 
CH$_3$CN           & $2{\sf x}10^{15}$ & $1{\sf x}10^{16}$       & --    & $8{\sf x}10^{16}\,^d$ \\
\hline
\hline
\end{tabular}
\footnotesize{~\\
  $^a$ Calculated for lower average $T$ of 200\,K because of smoothing to 5700\,AU resolution (Figs.~\ref{sample_spectra} \& \ref{ch3oh_sample}). The source size was approximated by half the spatial resolution. The on average lower Orion-KL column densities are likely due to the largest amount of missing flux for the closest source of the sample.\\
  $^b$ From \citet{beuther2007d}. \\
  $^c$ From \citet{beuther2007f} for different sub-sources.\\
  $^d$ From \citet{leurini2007} for different sub-sources.\\
  $^e$ At lower temperature of 100\,K, because otherwise different lines would get excited.
}
\label{column}
\end{table}

\end{document}